\documentclass[pdflatex,sn-mathphys-num]{sn-jnl}

\usepackage{graphicx}%
\usepackage{multirow}%
\usepackage{amsmath,amssymb,amsfonts}%
\usepackage{amsthm}%
\usepackage{mathrsfs}%
\usepackage[title]{appendix}%
\usepackage{xcolor}%
\usepackage{textcomp}%
\usepackage{manyfoot}%
\usepackage{booktabs}%
\usepackage{algorithm}%
\usepackage{algorithmicx}%
\usepackage{algpseudocode}%
\usepackage{listings}%
\usepackage{placeins}
\usepackage{float}   

\usepackage{pgfplots}
\pgfplotsset{compat=1.18}
\usepgfplotslibrary{groupplots,dateplot}
\usepackage{tikz}
\usetikzlibrary{arrows.meta,positioning,calc,fit,shapes,decorations.pathreplacing}

\definecolor{cBlue}{RGB}{40,90,170}
\colorlet{cGray}{black!55}
\colorlet{cLight}{black!8}
\colorlet{cGrid}{black!22}
\tikzset{
  arch font/.style={font=\footnotesize},
  arch box/.style={draw, rounded corners=2pt, line width=0.8pt, align=center,
                   inner sep=3.5pt, outer sep=0pt, minimum height=10mm, fill=white,
                   font=\footnotesize},
  arch arrow/.style={-{Stealth[length=2.3mm,width=2.0mm]}, line width=0.8pt},
  arch group/.style={draw, rounded corners=3pt, line width=0.6pt, fill=cLight},
  arch caption/.style={font=\footnotesize\sffamily},
}
\pgfplotsset{
  finance base/.style={
    font=\footnotesize,
    grid=major,
    grid style={cGrid},
    tick label style={font=\footnotesize},
    label style={font=\footnotesize},
    legend style={font=\footnotesize, draw=none, fill=none},
  },
}

\NewDocumentCommand{\pnlchart}{m m m m m O{-0.75} O{-0.75}}{%
\begin{figure}[H]\centering
\begin{tikzpicture}
\begin{groupplot}[
  group style={group size=1 by 4, vertical sep=10mm, x descriptions at=edge bottom},
  finance base,
  width=0.92\linewidth,
  date coordinates in=x,
  xticklabel={\year},
  xmin=2014-01-01, xmax=2025-03-01,
  enlarge x limits=false,
  tick align=outside,
  yticklabel style={text width=1.0cm, align=right},
]
\nextgroupplot[
  height=5.6cm,
  ylabel={Cumulative return},
  legend columns=2,
  legend to name=leg:#5,
  legend style={column sep=6pt},
]
\addplot[black, line width=0.9pt]
  table[col sep=comma, x=date, y=azb]{#1}; \addlegendentry{AlphaZeroBeta}
\addplot[cBlue, line width=0.9pt]
  table[col sep=comma, x=date, y=index]{#1}; \addlegendentry{Index buy-and-hold}
\addplot[cGray, dashed, line width=0.9pt]
  table[col sep=comma, x=date, y=mxsharpe]{#1}; \addlegendentry{MxSharpe baseline}
\addplot[cGray, densely dotted, line width=0.9pt]
  table[col sep=comma, x=date, y=decorr]{#1}; \addlegendentry{Decorr baseline}
\nextgroupplot[
  height=3.0cm,
  ylabel={Drawdown},
  ymin=#6, ymax=0.05,
  yticklabel={$\pgfmathprintnumber{\tick}$},
]
\addplot[black!30, line width=0.5pt, fill=cLight]
  table[col sep=comma, x=date, y=azb]{#2} \closedcycle;
\addplot[cBlue, dashed, line width=0.9pt]
  table[col sep=comma, x=date, y=index]{#2};
\nextgroupplot[
  height=3.0cm,
  ylabel={Long / short},
  ymin=#7, ymax=1.1,
  ytick={-0.5,0,0.5,1.0},
]
\addplot[cBlue!45, line width=0.5pt, fill=cBlue!18]
  table[col sep=comma, x=date, y=long]{#3} \closedcycle;
\addplot[cGray!45, line width=0.5pt, fill=cGray!18]
  table[col sep=comma, x=date, y=short]{#3} \closedcycle;
\draw[cGrid, dashed, line width=0.5pt]
  (axis cs:2014-01-01,0) -- (axis cs:2025-03-01,0);
\nextgroupplot[
  height=2.4cm,
  ylabel={Gross},
  xlabel={Year},
  ymin=0, ymax=1.25,
  ytick={0,0.5,1.0},
  skip coords between index={0}{0},
]
\addplot[black, line width=0.9pt]
  table[col sep=comma, x=date, y=gross]{#3};
\end{groupplot}
\node[above=6mm] at (group c1r1.north) {\pgfplotslegendfromname{leg:#5}};
\end{tikzpicture}
\caption{#4}\label{#5}
\end{figure}%
}

\newcounter{listing}[subsection]
\renewcommand{\thelisting}{\thesubsection.\arabic{listing}}
\newcommand{\listingcaption}[2]{%
  \refstepcounter{listing}%
  \label{#1}%
  \noindent\textbf{Listing \thelisting.} #2\par\vspace{0.4em}%
}

\theoremstyle{thmstyleone}%
\theoremstyle{thmstyletwo}%

\theoremstyle{thmstylethree}%

\begin{document}
\hypersetup{hypertexnames=false}

\title[AlphaZeroBeta: Deep Reinforcement Learning for Market-Neutral Portfolios]{AlphaZeroBeta: Deep Reinforcement Learning for Market-Neutral Portfolios}


\author*[1]{\fnm{Boris} \sur{Belyakov}}\email{work.belyakov@gmail.com}

\affil*[1]{\orgdiv{Department of Information Technology and Automated Systems, MIEM Faculty}, \orgname{HSE University}, \orgaddress{\street{11 Pokrovsky Bulvar}, \city{Moscow}, \postcode{109028}, \country{Russia}}}

\abstract{Market-neutral portfolios aim to generate consistent returns while offsetting systematic market risk. Traditional approaches based on factor models or convex optimization often underperform during market regime shifts or when structural assumptions break down. We propose AlphaZeroBeta, a deep reinforcement learning framework designed to deliver benchmark-relative alpha (excess returns) with near-zero beta (market neutrality). AlphaZeroBeta combines a composite reward function that balances risk-adjusted excess return, benchmark correlation, and transaction costs with a CNN-GRU policy trained end-to-end via Recurrent PPO and evaluated through a rolling walk-forward protocol. Backtests covering 2014--2024 across seven equity indices show that the model achieves higher Sharpe ratios than the baselines while maintaining near-zero benchmark correlations and competitive drawdowns.}




\maketitle

\vspace{35mm}

\section{Introduction}\label{sec:introduction}

Financial markets are inherently noisy and unpredictable, yet exhibit persistent patterns that can be systematically exploited using quantitative techniques. The efficient market hypothesis (EMH) posits that prices fully reflect available information~\cite{fama1970}; however, decades of empirical evidence reveal predictable anomalies such as momentum~\cite{jegadeesh1993}, excess volatility~\cite{shiller1981}, and multifactor risk premia~\cite{fama1993}. Market-neutral strategies, designed to isolate alpha (excess returns) while minimizing beta (systematic risk), have therefore emerged as a cornerstone of hedge funds and quantitative trading desks~\cite{patton2009, aqr2015}.

Despite their popularity, traditional market-neutral approaches face significant methodological and practical challenges:

\begin{enumerate}

\item Conventional implementations rely on static factor structures, such as the Fama--French family of models~\cite{fama1993, fama2015}, where exposures are assumed constant. However, factors themselves can become obsolete or attenuated as market regimes evolve, highlighting the limitations of static models and motivating dynamic alternatives~\cite{barigozzi2024}.

\item Statistical arbitrage models, particularly those employing high-dimensional regressions or machine learning, are prone to overfitting—capturing noise rather than signal—especially when historical data are limited. This problem has been documented in both factor-based covariance estimation~\cite{anderson2024} and machine-learning-driven approaches to asset pricing~\cite{ye2024}.

\item Many strategies lack explicit mechanisms to adapt to regime shifts, such as volatility clustering or macroeconomic shocks. Despite well-documented caveats of simple factor timing strategies~\cite{blitz2025}, adaptive portfolio frameworks~\cite{li2023} show that dynamic adjustments, incorporating time-varying predictors, regime indicators, or tactical factor allocation, can improve resilience compared to static models.

\end{enumerate}

Deep reinforcement learning (DRL) combines sequential decision-making with function approximation, providing an alternative to static, factor-based approaches. By casting portfolio construction as a Markov decision process (MDP), DRL learns state-contingent rebalancing policies that optimize intertemporal objectives; in practice, this supports dynamic risk management under frictions (e.g.\ transaction costs, exposure limits, and drawdown control) while exploiting time-varying structure in returns through representation learning over rolling training windows~\cite{sutton2018, jiang2017, buehler2019, spooner2018, almahdi2017}.

This study advances the literature through three principal contributions.

\begin{enumerate}
\item Introduction of AlphaZeroBeta, a deep reinforcement learning framework for the construction and management of market-neutral portfolios.
\item Proposal of a dynamic, risk-aware reward function and an explicit dollar-neutral projection that jointly enforce market neutrality while targeting benchmark-relative excess returns.
\item Provision of an empirical evaluation, employing historical equity data and benchmarking against both market indices and optimization models.
\end{enumerate}

Throughout this manuscript, ``alpha'' denotes residual returns relative to each benchmark index after accounting for standard risk factors; the model is not intended to capture a universal or economy-wide anomaly.

The remainder of the paper proceeds as follows. Section~\ref{sec:litreview} reviews the related literature on market-neutral portfolios and reinforcement learning for trading. Section~\ref{sec:methodology} details the architecture, reward, and MDP formulation. Section~\ref{sec:experiment} describes the data and baselines. Section~\ref{sec:compute} reports compute requirements, evaluation metrics, and the walk-forward protocol. Section~\ref{sec:results} presents empirical findings, robustness checks, and attribution analyses. Section~\ref{sec:discussion} discusses implications and future work; Section~\ref{sec:conclusions} concludes.

\section{Literature Review}\label{sec:litreview}

\subsection{Market-Neutral Portfolios}

Market-neutral investing is an established field in quantitative finance, with theoretical foundations, practical implementations, and empirical results spanning several decades. The original hedge fund, launched by Alfred W.\ Jones in 1949, sought to profit from stock selection while neutralizing market risk by combining long positions financed by short sales, with the long and short legs sized to offset overall market exposure. \citet{patton2009} documents that standard industry definitions of equity market-neutral strategies center on this long--short construction for neutralizing systematic risks, and argues that these definitions are inadequate for the nonlinear payoffs typical of hedge funds; he therefore proposes richer notions of neutrality (mean, variance, value-at-risk, and tail).

The conceptual basis for market-neutral investing is grounded in multi-factor asset-pricing models. Fama and French~\cite{fama1993} introduced a three-factor model showing that stock returns are explained by market excess return, size, and value factors. Carhart~\cite{carhart1997} extended this with a momentum factor defined by a self-financing portfolio long high-momentum stocks and short low-momentum stocks. The factor-mimicking portfolios are structured in long--short form, and hedging their associated exposures yields zero-beta portfolios.

From a methodological perspective, classical approaches include regression-based factor-neutralization~\cite{avellaneda2010} and optimization techniques imposing zero-exposure constraints~\cite{markowitz1952, ganesan2011}. Relative-value strategies, most notably pairs trading, represent standard market-neutral approaches. Gatev, Goetzmann, and Rouwenhorst (2006) demonstrate that simple self-financing pairs trading strategies implemented on U.S. equities over the period 1962–2002 yield statistically and economically significant performance, with average annualized excess returns on the order of 11\%~\cite{gatev2006}. Convertible bond arbitrage and merger arbitrage are other widely studied market-neutral techniques~\cite{patton2009, hedgenordic2015}.

Over the past decade, machine learning techniques have substantially broadened the methodological basis for statistical arbitrage. Krauss, Do, and Huck (2017) showed that deep neural networks, gradient-boosted trees, and random forests can systematically generate profitable long--short portfolios on the S\&P~500~\cite{krauss2017}. Subsequent work extends this trend toward end-to-end deep learning for portfolio optimization~\cite{zhang2020a} and reinforcement learning for trading~\cite{zhang2020b}. The August 2007 quant deleveraging episode demonstrated that market-neutral books are not free of systematic risk in stressed regimes~\cite{khandani2007, khandani2011}, motivating an explicit, correlation-based neutrality penalty embedded directly in the reward function (Section~\ref{subsec:reward}) rather than reliance on factor cancellation alone.

From a practical standpoint, a classical approach to constructing market-neutral portfolios relies on factor models to disentangle asset returns from broader market influences. For example, Avellaneda and Lee (2010) introduced a method that involves regressing individual stock returns on multiple factors, such as sector-specific and market-wide indices~\cite{avellaneda2010}. The goal of this methodology is to construct portfolios with zero net exposure to these factors, thereby achieving market neutrality. This regression-based framework reduces the complexity of portfolio construction by expressing asset returns through their factor loadings and enabling the development of portfolios that minimize systematic risk exposure.

The mean-variance optimization framework introduced by Markowitz (1952) has also been adapted for constructing market-neutral portfolios~\cite{markowitz1952}. For example, Ganesan (2011) applied this approach while imposing constraints to ensure zero exposure to systematic risk factors, using a subspace-decomposition analysis on a broad U.S. equity universe (roughly 2{,}000 NYSE/NASDAQ stocks, 1998--2010)~\cite{ganesan2011}. The constraints kept the resulting portfolios market-neutral, demonstrating that optimization-based methods can sustain neutrality on large universes.

A notable market-neutral strategy variation is pairs trading, where a portfolio consists of two assets: one held long and the other short. The goal is to exploit relative price movements between the two, often under the assumption that their prices will converge or diverge in a predictable manner. Baronyan et al.\ (2010) compared fourteen pairs-trading strategies (combinations of seven trading rules and two pair-selection methods) and found that performance varies materially across configurations and regimes~\cite{baronyan2010}. This dependence on strategy specification underscores that market neutrality admits multiple operationalizations and that the chosen criterion materially influences portfolio outcomes.

Valle et al.\ (2014) propose a Mixed-Integer Nonlinear Program (MINLP) that constructs market-neutral portfolios by minimizing the absolute correlation between portfolio returns and a benchmark index, allowing both long and short positions~\cite{valle2014market}. Their formulation attains low (often near-zero) correlations and, evaluated against S\&P international universes, outperforms both the underlying indices and a zero-beta baseline (minimizing the regression slope), as well as several existing market-neutral funds.

Deep reinforcement learning, while still emerging in financial research, offers a framework for constructing market-neutral portfolios. The ability of DRL to learn optimal policies from market data makes it an attractive approach for portfolio management. Unlike traditional regression-based methods, DRL algorithms can learn to optimize both risk and return without requiring explicit factor models. The growing body of literature on DRL for finance has demonstrated the potential of DRL for investment decision-making~\cite{moody2001}. However, the application of DRL to market-neutral portfolios remains relatively unexplored.

Our study advances the field of market-neutral portfolio construction by introducing a DRL framework that bypasses traditional beta estimation and regression-based techniques. Instead, the agent learns to directly optimize portfolio weights while explicitly projecting them to dollar-neutral at every step and using a correlation-aware reward to discourage residual beta.

Building on prior research that relied on limited trading settings or asset universes, our method scales to portfolios with 500+ stocks by employing DRL techniques such as Recurrent Proximal Policy Optimization (Recurrent PPO) with Gated Recurrent Units (GRUs). This supports a dynamic, data-driven investing process that adjusts to changing market conditions without manual rebalancing or explicit factor modeling.

\subsection{Reinforcement Learning in Trading}

Reinforcement learning (RL) has been increasingly applied to finance, where it learns adaptively from sequential market interactions. Unlike traditional econometric and statistical methods (e.g.\ SARIMAX, VAR), RL frameworks combine online learning with deep network architectures to handle the noise, non-stationarity, heavy-tailed distributions, and multi-participant dynamics that characterize financial data. As reviewed by Charpentier et al.\ (2023) and Huang et al.\ (2020), these methods continuously update policies as new data arrives, which suits them to markets that exhibit volatility clustering and abrupt regime shifts~\cite{charpentier2020, huang2020}.

The reinforcement learning literature broadly categorizes methods into model-free approaches, including actor-only, critic-only, and actor-critic frameworks. Comparative studies reviewed by Fischer (2018) report no significant performance differences between core algorithms such as Policy Gradient (PG) and Deep Q-Network (DQN) when applied to portfolio management~\cite{fischer2018}. However, each class of methods offers specific strengths depending on the task setting~\cite{pricope2021}. MDP design choices materially affect agent performance. For example, augmenting a DRL portfolio allocator with market-sentiment features yields significantly higher Sharpe ratios and annualized returns than price-only baselines~\cite{koratamaddi2021}.

Evaluations in market making and portfolio optimization, as highlighted by Gašperov et al.\ (2021) and Hambly et al.\ (2023), consistently show that RL methods can manage larger action spaces more efficiently than manual optimization~\cite{gasperov2021, hambly2023}. Enhanced state representations tend to improve learning outcomes, though diminishing returns are observed as feature space expands. In multi-asset scenarios, an expanded action space allows finer control over asset allocation, reducing the burden on human analysts. While most RL implementations continue to use portfolio return as the primary reward signal, recent research has introduced alternative objectives—such as risk-adjusted returns and multi-objective reward shaping—to better align with real-world portfolio management goals~\cite{bai2024, choudhary2025, orra2025}.

Simplified trading assumptions—such as neglecting transaction costs or assuming zero slippage—risk overestimating RL performance, emphasizing the need for rigorous models that incorporate these constraints~\cite{bai2024}.

AlphaZeroBeta addresses these challenges through a risk-aware reward function and dynamic policy adaptation tailored to the non-stationary structure of financial data. State representations combine volatility-adjusted features with cross-sectional normalization across sector and index peers, allowing the model to learn from heavy-tailed distributions and long-range dependencies without overfitting to transient market noise.

\subsection{State-of-the-Art Neural Architectures}

Choosing the recurrent backbone is crucial because market-neutral trading involves long, non-stationary sequences with sparse but impactful regime shifts. GRUs~\cite{cho2014} share the gating mechanism of long short-term memory (LSTM) networks but use fewer parameters and have been competitive on time-series tasks. Transformer variants have been adapted to long-range memory (Transformer-XL~\cite{dai2019}) and to reinforcement learning specifically (Gated Transformer-XL~\cite{parisotto2020}), but Pleines et al.\ (2025)~\cite{pleines2025} show that GRU-based policies maintain stable performance on endless reinforcement learning tasks, whereas transformer variants suffer when memory must be updated continuously. Empirically, we observe the same behavior on financial data: GRUs handle the rolling multi-year training horizons used in our walk-forward protocol (see Section~\ref{sec:walk_forward_validation}) with fewer parameters and more stable optimization, so we adopt a CNN-GRU encoder instead of a transformer stack.

\section{Methodology}\label{sec:methodology}

\begin{figure}[htbp]
\centering
\begin{tikzpicture}[
  font=\footnotesize,
  card/.style={draw=black!70, line width=0.4pt, fill=white, rounded corners=1pt,
               align=center, font=\footnotesize, inner sep=1.5pt,
               minimum width=27mm, minimum height=11mm},
  cardback/.style={draw=black!55, line width=0.3pt, fill=white, rounded corners=1pt,
                   minimum width=27mm, minimum height=11mm},
  cube/.style={draw=black!70, line width=0.45pt, fill=cBlue!10,
               rounded corners=1pt, align=center, font=\footnotesize,
               minimum width=24mm, minimum height=14mm, inner sep=2pt},
  listbox/.style={draw=black!70, line width=0.45pt, fill=white, rounded corners=1pt,
                  align=left, font=\footnotesize, text width=46mm, inner sep=4pt},
  pill/.style={draw=black!70, rounded corners=2pt, line width=0.45pt, fill=white,
               align=center, font=\footnotesize, inner sep=2pt,
               minimum width=30mm, minimum height=9mm},
  agent/.style={draw=cBlue!85, dashed, rounded corners=1.5pt, line width=0.5pt,
                fill=white, align=center, font=\footnotesize,
                minimum width=30mm, minimum height=11mm},
  cyl/.style={draw=black!70, cylinder, shape border rotate=90, aspect=0.22,
              line width=0.45pt, fill=cBlue!10, align=center, font=\footnotesize,
              minimum width=28mm, minimum height=14mm},
  section/.style={draw=black!55, dashed, line width=0.5pt, rounded corners=3pt},
  stagetitle/.style={font=\footnotesize\bfseries, anchor=west,
                     fill=white, inner xsep=4pt, inner ysep=1pt, text=black!80},
  flow/.style={-{Stealth[length=2.2mm,width=1.8mm]}, line width=0.5pt, draw=black!80},
  flowlabel/.style={midway, fill=white, inner sep=1.5pt, font=\footnotesize, align=center},
  bigflow/.style={-{Stealth[length=2.6mm,width=2.0mm]}, line width=0.6pt, draw=black!80},
  obstext/.style={font=\footnotesize, align=right, inner sep=1pt, text=black!85},
  arrowlabel/.style={font=\footnotesize, align=center, fill=white, inner sep=1pt},
]

\foreach \x/\y/\name in {%
   -36/68/{Stock\\Prices}, -8/68/{Company\\Profiles}, 20/68/{Financial\\Statements}, 48/68/{Commodity\\Prices},
   -36/55/{Stock\\Volumes}, -8/55/{Dividends\\\& Splits}, 20/55/{Statement\\Analysis}, 48/55/{Forex\\Prices},
   -36/42/{Social\\Sentiments}, -8/42/{Earnings\\Surprises}, 20/42/{Economics\\Data}, 48/42/{Index\\Values}%
} {
   \node[cardback] at (\x mm + 2.0mm, \y mm - 1.3mm) {};
   \node[cardback] at (\x mm + 1.0mm, \y mm - 0.65mm) {};
   \node[card]     at (\x mm,         \y mm) {\name};
}
\node[section, fit={(-54mm, 35mm) (66mm, 82mm)}] (s1) {};
\node[stagetitle] at (-52mm, 82mm) {Data Loading};

\draw[bigflow] (8mm, 35mm) -- (8mm, 31mm);
\node[font=\footnotesize\itshape, anchor=west, fill=white, inner xsep=2pt, inner ysep=0pt]
  at (10mm, 33mm) {Processing};

\node[cube]    (agg)       at (-36mm, 12mm) {Aggregated\\Data};
\node[listbox] (alphalist) at (8mm, 12mm) {%
  1.\;Momentum (pct\_change)\\
  2.\;Mean Reversion (-1*mean)\\
  3.\;Volatility (std)\\
  4.\;Skew / Jump (skew, kurtosis)%
};
\node[cube]    (tensor)    at (48mm, 12mm) {Input\\Tensor};

\draw[flow] (agg.east) -- (alphalist.west);
\draw[flow] (alphalist.east) -- (tensor.west);
\node[arrowlabel]
  at ($(agg.east)!0.5!(alphalist.west) + (0, 12mm)$) {Derive alpha\\signals};
\node[arrowlabel]
  at ($(alphalist.east)!0.5!(tensor.west) + (0, 12mm)$) {Normalize across\\sectors / index};

\node[section, fit={(-54mm, -1mm) (66mm, 30mm)}] (s2) {};
\node[stagetitle] at (-52mm, 30mm) {Feature Engineering};

\draw[bigflow] (8mm, -1mm) -- (8mm, -5mm);
\node[font=\footnotesize\itshape, anchor=west, fill=white, inner xsep=2pt, inner ysep=0pt]
  at (10mm, -3mm) {Modeling};

\node[pill] (feed) at (-36mm, -13mm) {Simulated\\Live Feed};
\node[anchor=west, font=\footnotesize, align=left, text width=78mm] (cap)
  at (-18mm, -13mm) {Rolling 36-month training window, 6-month test slice, daily rebalancing, retrained per fold.};
\draw[flow] (feed.east) -- (-18mm, -13mm);
\draw[flow] (8mm, -17mm) -- (8mm, -31.5mm);

\node[pill] (ppo) at (48mm, -25mm) {RecurrentPPO\\(CnnGRU)};

\node[agent] (agent) at (8mm, -37mm) {Trading Agent};
\node[agent] (env)   at (8mm, -55mm) {Environment};

\draw[flow] (ppo.south) |- (agent.east);

\node[obstext, anchor=east] (obs) at (-15mm, -37mm) {Observation, $O_t$};
\draw[flow] (obs.east) -- (agent.west);

\coordinate (act_top) at ($(agent.south) + (9mm, 0)$);
\coordinate (act_bot) at ($(env.north)  + (9mm, 0)$);
\draw[flow] (act_top) -- (act_bot);
\node[anchor=west, font=\footnotesize]
  at ($(act_top)!0.5!(act_bot) + (2mm, 0)$) {Action, $A_t$};

\coordinate (rew_bot) at ($(env.north)  + (-9mm, 0)$);
\coordinate (rew_top) at ($(agent.south) + (-9mm, 0)$);
\draw[flow] (rew_bot) -- (rew_top);
\node[anchor=east, font=\footnotesize]
  at ($(rew_bot)!0.5!(rew_top) + (-2mm, 0)$) {Reward, $R_t$};

\node[cyl] (orders)  at (-36mm, -68mm) {Market Orders\\$+$ Slippage};
\node[cyl] (results) at (48mm, -68mm) {Trading\\Results};
\draw[flow, rounded corners=2.5pt] (env.west) -| (orders.north);
\draw[flow] (orders.east) --
  node[flowlabel]{Daily Portfolio\\Rebalancing $+$ Commission} (results.west);

\node[section, fit={(-54mm, -76mm) (66mm, -5mm)}] (s3) {};
\node[stagetitle] at (-52mm, -5mm) {Backtesting};

\end{tikzpicture}
\caption{AlphaZeroBeta architecture: a three-stage pipeline. \textbf{Data Loading} aggregates twelve heterogeneous market and alternative data sources. \textbf{Feature Engineering} derives four alpha-signal families (momentum, mean reversion, volatility, and skew/jump) over rolling windows and normalizes them per sector and index into the input tensor. \textbf{Backtesting} formalizes the MDP loop: a Recurrent PPO trading agent with a CNN-GRU encoder ingests the observation (tensor slice, historical weights, portfolio metrics), emits an action (asset weights), and receives a reward (the AlphaZeroBeta composite reward) from a transaction-cost-aware environment.}\label{fig:alpha_zero_beta_arch}
\end{figure}
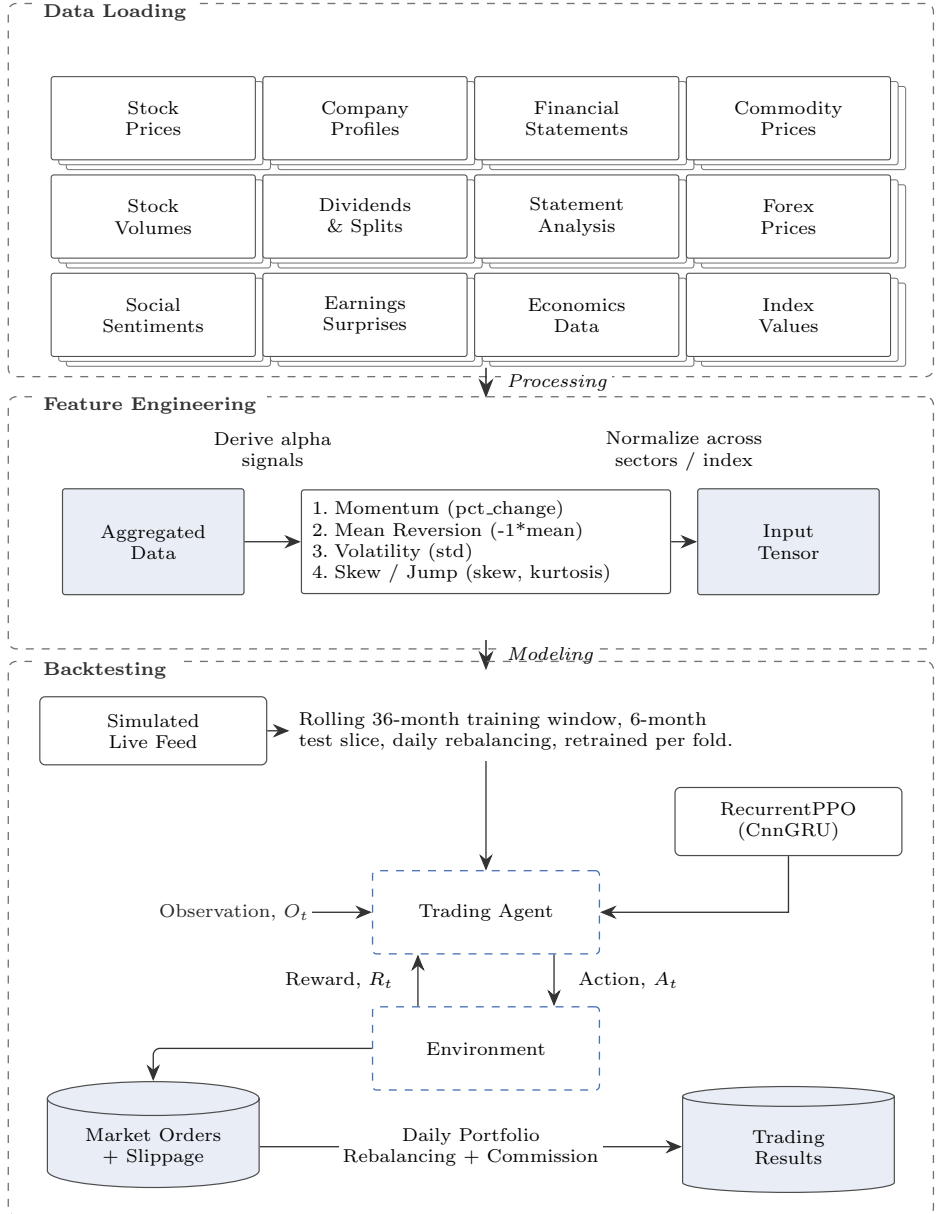

The AlphaZeroBeta architecture is structured around three core stages to achieve dynamic market-neutral portfolio optimization. An overview of the full system architecture is shown in Figure~\ref{fig:alpha_zero_beta_arch}.

\subsection{Alpha Signal Construction}

In the first stage, heterogeneous financial data sources—including price histories, fundamental metrics, and sentiment indicators—are aggregated and transformed into a standardized set of alpha signals. These signals are engineered to capture a diverse range of market phenomena, including momentum, mean reversion, volatility clustering, and tail risk asymmetries. The resulting features are arranged into a multi-dimensional tensor that preserves both temporal and cross-sectional structure, serving as a unified input representation for the model.

\subsection{Feature Extraction via CNN-GRU}

The constructed signal tensor is then passed through a hybrid feature extractor composed of a CNN followed by a GRU\@. The CNN component extracts multi-scale temporal features through three 1D convolutional layers with hierarchical receptive fields. The GRU component then captures longer-range temporal dependencies and retains a compact memory of historical context. Together, these layers model complex non-linear patterns and stabilize performance across market regimes.

\subsection{Policy Learning with Recurrent PPO}

Extracted features are forwarded to a Recurrent PPO agent, an actor-critic algorithm based on PPO, extended for recurrent policies~\cite{schulman2017, pleines2022, sb3contrib2025}. The agent outputs portfolio weights by optimizing a composite reward function designed to favor excess returns per unit of volatility while penalizing correlation with a benchmark index and transaction costs. Formally, at each decision step the PPO agent receives the scalar reward $r_t$ defined in Eq.~\eqref{eq:reward_full} in Subsection~\ref{subsec:reward}, which combines risk-adjusted excess return with explicit penalties for market correlation and turnover. This reward is used directly as the optimization objective in the policy gradient updates.

\subsection{Walk-Forward Evaluation}

To assess the model’s generalization and robustness, we run walk-forward backtests over the 2014--2024 period, reserving earlier years for training context. At the start of each split, we train the agent on a fixed-length historical window; we then deploy it on the subsequent out-of-sample period, ensuring non-overlapping train/test segments within each split.

The procedure simulates daily trading operations and allows the model to adapt to regime shifts while limiting overfitting, providing validation under realistic, non-stationary market conditions.

\subsection{Problem Formulation as a Markov Decision Process}

We formalize market-neutral portfolio construction as a Markov decision process, defined by the tuple $(\mathcal{S}, \mathcal{A}, \mathcal{P}, \mathcal{R}, \gamma)$, where $\mathcal{S}$ denotes the state space, $\mathcal{A}$ the action space, $\mathcal{P}$ the transition dynamics, $\mathcal{R}$ the reward function, and $\gamma \in [0,1]$ the discount factor~\cite{sutton2018}.

At time $t$, the environment is characterized by a state $s_t \in \mathcal{S}$, which includes the current portfolio weights $\mathbf{w}_t$, global market signals $\mathbf{m}_t$ (macroeconomic indicators, commodity prices, and broad market indices), and asset-specific features $\mathbf{x}_t$ (price history, fundamentals, and sentiment). A detailed taxonomy is provided in Appendix~\ref{app:features}. Formally,
\begin{equation}
s_t = (\mathbf{w}_t, \mathbf{m}_t, \mathbf{x}_t).
\end{equation}

The agent selects an action $a_t \in \mathcal{A}$ corresponding to the target allocation $\mathbf{w}_{t+1}$ for the next period. To keep leverage bounded we restrict the weights to lie in an $\ell_1$ ball, $\|\mathbf{w}_{t+1}\|_1 \leq \tau$, with $\tau=1$ in every experiment, and we subtract the cross-sectional mean of the action so that $\sum_i w_i = 0$ before the $\ell_1$ projection (see Listing~\ref{lst:alphazerobeta-env} in Appendix~\ref{app:extra-code}). This combination enforces dollar neutrality at each rebalance while keeping gross exposure at most one. The state transition to $s_{t+1}$ follows a stochastic process governed by market evolution and rebalancing:
\begin{equation}
s_{t+1} \sim \mathcal{P}(\cdot \mid s_t, a_t).
\end{equation}

The reward function $\mathcal{R}(s_t, a_t)$ captures the agent's objective, which balances return generation with risk control and neutrality preservation.

The agent's goal is to learn a policy $\pi: \mathcal{S} \rightarrow \mathcal{A}$ that maximizes the expected discounted cumulative reward:
\begin{equation}
\pi^* = \arg\max_{\pi} \; \mathbb{E}_\pi \left[ \sum_{t=0}^{T} \gamma^t r_t \right],
\end{equation}
where $r_t = \mathcal{R}(s_t, a_t)$ denotes the immediate reward at time $t$~\cite{sutton2018}.

\subsection{Reinforcement Learning Architecture}

To optimize the policy $\pi_\theta(a_t \mid s_t)$ in a Markov decision process, we employ Recurrent PPO, an actor-critic algorithm characterized by stability and sample efficiency in continuous control tasks~\cite{schulman2017, pleines2022}. The policy is parameterized by a deep neural architecture that processes raw market data and outputs portfolio allocations in a closed-loop, state-dependent setting.

The input to the agent consists of multivariate financial time series, sampled at multiple temporal resolutions—specifically, daily, weekly, and monthly intervals—to encode both short-term fluctuations and long-term structural trends. Weekly and monthly tensors are constructed by resampling the daily panel using end-of-period values so that every observation shares the timestamp of the most recent daily close, ensuring synchronous alignment when the streams are concatenated. The multi-scale observations are then processed by a hierarchical convolutional pipeline, where three 1D convolutional layers apply progressively narrower filters of sizes 8, 4, and 3, with corresponding strides of 4, 2, and 1. The architecture extracts multi-scale temporal features across assets and indicators.

The resulting feature map is then flattened and passed through a GRU with 512 hidden units. The GRU maintains temporal dependencies and encodes sequential patterns, providing a memory mechanism for evolving market dynamics.

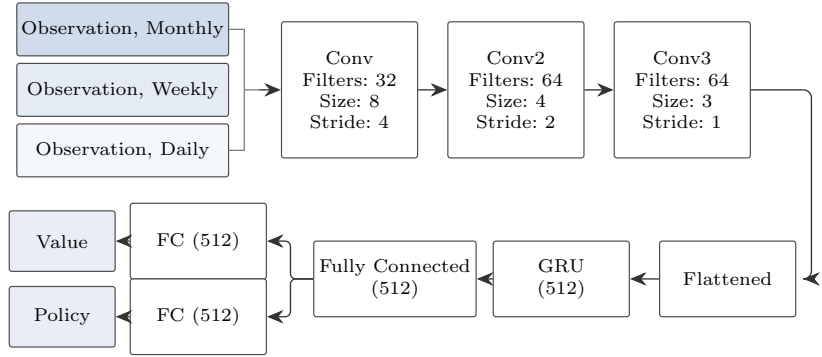
\begin{figure}[htbp]
\centering
\begin{tikzpicture}[
  font=\footnotesize,
  obs/.style={draw=black!70, line width=0.45pt, fill=#1, rounded corners=1pt,
              minimum width=28mm, minimum height=7mm,
              align=center, font=\footnotesize, inner sep=1.5pt},
  conv/.style={draw=black!70, line width=0.45pt, fill=white, rounded corners=1pt,
               minimum width=18mm, minimum height=18mm,
               align=center, font=\footnotesize, inner sep=2pt},
  layer/.style={draw=black!70, line width=0.45pt, fill=white, rounded corners=1pt,
                minimum width=18mm, minimum height=10mm,
                align=center, font=\footnotesize, inner sep=2pt},
  head/.style={draw=black!70, line width=0.45pt, fill=cBlue!10, rounded corners=1pt,
               minimum width=14mm, minimum height=8mm,
               font=\footnotesize, align=center},
  flow/.style={-{Stealth[length=2.2mm,width=1.8mm]}, line width=0.5pt, draw=black!80},
  bracket/.style={line width=0.45pt, draw=black!55},
]
\node[obs=cBlue!22] (m) at (14mm,  11mm) {Observation, Monthly};
\node[obs=cBlue!12] (w) at (14mm,   3mm) {Observation, Weekly};
\node[obs=cBlue!5]  (d) at (14mm,  -5mm) {Observation, Daily};

\coordinate (in_e) at (32mm, 3mm);
\draw[bracket] (m.east) -- ([xshift=2mm]m.east) -- ([xshift=2mm]d.east) -- (d.east);
\draw[bracket] ([xshift=2mm]w.east) -- (in_e);

\node[anchor=south, font=\footnotesize] at (60mm, 22mm)
  {Stacked input: $3 \times n_{\text{features}} \times n_{\text{agent\_window}}$};

\node[conv] (c1) at (44mm, 3mm) {Conv\\Filters: 32\\Size: 8\\Stride: 4};
\node[conv] (c2) at (66mm, 3mm) {Conv2\\Filters: 64\\Size: 4\\Stride: 2};
\node[conv] (c3) at (88mm, 3mm) {Conv3\\Filters: 64\\Size: 3\\Stride: 1};

\draw[flow] (in_e) -- (c1.west);
\draw[flow] (c1.east) -- (c2.west);
\draw[flow] (c2.east) -- (c3.west);

\draw[flow, rounded corners=3pt]
  (c3.east) -- ++(8mm,0) |- (104mm, -22mm);

\node[layer] (flat) at (94mm,  -22mm) {Flattened};
\node[layer] (gru)  at (72mm,  -22mm) {GRU\\(512)};
\node[layer] (fc0)  at (50mm,  -22mm) {Fully Connected\\(512)};

\node[layer] (fcv) at (24mm,  -17mm) {FC (512)};
\node[layer] (fcp) at (24mm,  -27mm) {FC (512)};

\node[head] (val) at (6mm,  -17mm) {Value};
\node[head] (pol) at (6mm,  -27mm) {Policy};

\draw[flow] (flat.west) -- (gru.east);
\draw[flow] (gru.west)  -- (fc0.east);
\draw[flow, rounded corners=2pt] (fc0.west) -- ++(-3mm,0) |- (fcv.east);
\draw[flow, rounded corners=2pt] (fc0.west) -- ++(-3mm,0) |- (fcp.east);
\draw[flow] (fcv.west) -- (val.east);
\draw[flow] (fcp.west) -- (pol.east);
\end{tikzpicture}
\caption{AlphaZeroBeta neural architecture. Three multi-resolution observation tensors (daily, weekly, and monthly), each resampled at the same close, are stacked along the channel axis and processed by a sequential CNN (8/4/3 kernels, 4/2/1 strides, 32/64/64 filters). The flattened feature map feeds a GRU (512 hidden units) and a shared fully connected layer (512 units); the recurrent embedding then branches into a value head that outputs a scalar value estimate and a policy head that outputs an N-dimensional weight vector, where N is the universe size (e.g.\ 30 for the DJIA, 500 for the S\&P 500, about 2,200 for the SSE Composite). Each head has a 512-unit ReLU hidden layer; the policy output is bounded by a Tanh. Trained end-to-end with Recurrent PPO.}\label{fig:rl_architecture}
\end{figure}

The recurrent embedding is then forwarded through two separate fully connected branches. Because the action dimension equals the number of tradable assets in each universe, including cases with more than 2{,}200 constituents (000001.SS), the final policy layer is a high-dimensional linear projection from the 512-unit embedding to the per-asset weight vector. During training we batch experiences along both the time and asset axes (i.e.\ minibatches contain 32--64 parallel environments, each representing a different index window) so that GPU matrix multiplications dominate the cost rather than per-asset loops.

Inference remains tractable because the policy outputs a dense weight vector once per rebalance; sparsity is enforced afterward by projecting onto the $\ell_1$ ball, which naturally limits gross exposure without requiring explicit pruning. This design keeps runtime near-linear in the number of assets and allows the same PPO implementation to handle 500-name U.S. universes and the 2{,}200-name Shanghai panel with identical code.

The two fully connected branches are:
\begin{enumerate}
  \item a policy head that outputs the action vector $a_t = \mathbf{w}_{t+1}$, representing the next-period portfolio weights, and
  \item a value head that estimates $V_\phi(s_t)$, the expected return-to-go from state $s_t$.
\end{enumerate}

Each head contains a 512-unit fully connected ReLU hidden layer followed by a head-specific output layer: a scalar for $V_\phi(s_t)$ and an $N$-dimensional Tanh-bounded weight vector for $\mathbf{w}_{t+1}$, where $N$ is the universe size and varies by market (e.g.\ $30$ for \textasciicircum{}DJI, $500$ for \textasciicircum{}GSPC, ${\sim}2{,}200$ for 000001.SS). The entire network is trained end-to-end using PPO's clipped surrogate objective, which balances policy improvement with trust-region constraints. Gradient updates are computed via backpropagation using minibatch stochastic gradient descent. The overall architecture is illustrated in Figure~\ref{fig:rl_architecture}.

\subsection{Reward Function}\label{subsec:reward}

In reinforcement learning for portfolio optimization, the choice of reward function is critical: it determines which trade-offs the agent learns to make between return, risk, neutrality, and turnover. In AlphaZeroBeta we design the reward to encourage market-neutral excess returns with controlled volatility and realistic transaction costs, rather than pure return maximization.

Let $r_p(t)$ and $r_m(t)$ denote the portfolio and benchmark (index) returns at time $t$, respectively. We first define the excess return
\begin{equation}
\Delta r(t) = r_p(t) - r_m(t),
\end{equation}
which captures performance attributable to active decisions beyond passive index exposure. Because excess return in isolation does not account for risk, we normalize by the portfolio’s rolling volatility $\sigma_p(t)$ to obtain a Sharpe-like, risk-adjusted term
\begin{equation}
R_t^{(0)} = \frac{\Delta r(t)}{\sigma_p(t)}.
\end{equation}

To promote market neutrality, we penalize correlation between the portfolio and the benchmark. Let $\mathrm{Corr}\bigl(r_p(t), r_m(t)\bigr)$ denote the correlation between portfolio and benchmark returns, computed over a rolling window (e.g.\ 60 business days). We extend the reward to
\begin{equation}
R_t^{(1)} = \frac{\Delta r(t)}{\sigma_p(t)} \;-\; \lambda_1 \cdot \mathrm{Corr}\bigl(r_p(t), r_m(t)\bigr),
\end{equation}
where $\lambda_1 \ge 0$ is a regularization coefficient controlling the strength of the neutrality penalty. Larger values of $\lambda_1$ push the policy toward lower beta and weaker co-movement with the index.

Excessive turnover is undesirable both because it erodes returns via transaction costs and because it makes strategies harder to implement at scale. Let $w_i(t)$ denote the weight of asset $i$ at time~$t$. We approximate proportional transaction costs using
\begin{equation}
C_t = \sum_i \bigl|\Delta w_i(t)\bigr|, \qquad
\Delta w_i(t) = w_i(t) - w_i(t-1),
\end{equation}
which is proportional to portfolio turnover at each rebalance. This leads to the final reward
\begin{equation}
R_t = \frac{r_p(t) - r_m(t)}{\sigma_p(t)} \;-\; \lambda_1 \cdot \mathrm{Corr}\bigl(r_p(t), r_m(t)\bigr) \;-\; \lambda_2 \sum_i \bigl|\Delta w_i(t)\bigr|,
\label{eq:reward_full}
\end{equation}
where $\lambda_2 \ge 0$ controls the strength of the turnover penalty and can be calibrated to reflect typical cost levels in the underlying markets.

In the RL formulation, we take $r_t := R_t$ as the instantaneous reward in the Markov decision process. Thus the agent maximizes the expected discounted sum $\mathbb{E}\bigl[\sum_t \gamma^t R_t\bigr]$, trading off (i) risk-adjusted excess return, (ii) market correlation, and (iii) trading intensity. Coefficients $\lambda_1$ and $\lambda_2$ are treated as hyperparameters and kept fixed across markets in our main experiments; their values and other key PPO settings are summarized in Appendix~\ref{app:hyperparams}. By construction, this objective encourages policies that generate market-neutral excess returns while maintaining controlled volatility and realistic turnover.

Because the turnover penalty depends on the absolute change in weights, the environment becomes path-dependent: the reward earned at $t$ is influenced by the entire sequence of past allocations through transaction costs. Each fold therefore starts from a flat, self-financing portfolio, and the agent must ramp up exposures sequentially. In practice we observe that portfolio weights converge toward their stationary distribution after roughly ten trading days; this short warm-up period is included in all reported metrics, but it also explains the small dip in performance evident at the beginning of each out-of-sample window.

\section{Experiment Setup}\label{sec:experiment}

\subsection{Data}

We conduct empirical evaluation on a dataset retrieved from Bloomberg Terminal and Financial Modeling Prep, covering calendar years 2004--2024 of historical equity and macro-financial data~\cite{bloomberg2025, fmp2025}. The dataset comprises daily closing prices, trading volumes, and corporate actions (e.g.\ dividends, stock splits), alongside detailed company-level fundamentals such as income statements, balance sheets, financial ratios, and sector classifications. Additionally, we include analyst-estimated earnings surprise signals, which capture deviations between expected and reported earnings and serve as a proxy for information shocks. Walk-forward experiments draw their training data from rolling three-year windows (approximately 756 trading days), followed by a six-month validation segment (about 126 days) used for early stopping and hyperparameter checks, and a six-month out-of-sample test segment.

The block advances in six-month steps (sliding window): for the first fold, the model trains on July~2010--June~2013 (36 months), validates on July--December~2013, and tests on January--June~2014; the next fold shifts forward six months (train January~2011--December~2013, validate January--June~2014, test July--December~2014), and so on.

Because the six-month sliding step is shorter than the 36-month training window, training windows overlap across folds (by 30 months between consecutive folds); the six-month validation and test slices advance disjointly, yielding $K=22$ non-overlapping test segments through December~2024. The 2004--2010 segment is reserved for initialization of rolling estimators (volatility, correlation, and winsorization thresholds) and feature standardization statistics, so the first training window starts in July~2010 and no estimator state leaks across the in-sample/out-of-sample boundary. At the beginning of each fold, the agent's network weights are reset and retrained from scratch on that fold's training window; portfolio weights start from zero (flat position) and adapt dynamically during deployment. Unless otherwise noted, every descriptive statistic and inference in Sections~\ref{sec:experiment} and~\ref{sec:results} is computed on the 2014--2024 out-of-sample walk-forward window; the pre-2014 data only provide historical context for initializing each fold.

\begin{table}[htbp]
\small
\caption{Overview of equity indices used in the empirical study. Indices vary by region, coverage, and rebalancing conventions.}\label{tab:indices}
\begin{tabular*}{\textwidth}{@{\extracolsep\fill}lp{1.7cm}p{1.1cm}p{0.9cm}p{0.8cm}p{4.9cm}}
\toprule
\textbf{Code} & \textbf{Name} & \textbf{Region} & \textbf{Size} & \textbf{Curr.} & \textbf{Notes} \\
\midrule
000001.SS & SSE Composite & China & $>$2{,}200 & CNY & Comprehensive index tracking all A- and B-shares listed on the Shanghai Stock Exchange; used as a proxy for Chinese market breadth. \\
\textasciicircum{}DJI & Dow Jones Ind. Avg & U.S. & 30 & USD & Price-weighted index of 30 U.S. blue-chip companies; oldest and most concentrated benchmark in the study, with infrequent constituent changes. \\
\textasciicircum{}FTSE & FTSE 100 & U.K. & 100 & GBP & Tracks the 100 largest companies on the London Stock Exchange by market capitalization; rebalanced quarterly; sector-heavy in energy and financials. \\
\textasciicircum{}GDAXI & DAX & Germany & 40 & EUR & Comprises 40 major German companies trading on Xetra; free-float adjusted; used as the flagship index for the Eurozone’s largest economy. \\
\textasciicircum{}GSPC & S\&P 500 & U.S. & 500 & USD & Broad-based index of large-cap U.S. equities; represents ~80\% of total U.S. market cap; free-float weighted and rebalanced quarterly. \\
\textasciicircum{}HSI & Hang Seng & Hong Kong SAR, China & 83 & HKD & Captures the largest and most liquid companies on the Hong Kong Stock Exchange (Hong Kong SAR, China); rebalanced quarterly; includes dual-listed mainland firms. \\
\textasciicircum{}NDX & NASDAQ-100 & U.S. & 100 & USD & Tech-focused index of the largest non-financial NASDAQ-listed firms; uses modified capitalization weighting to reduce dominance of megacaps. \\
\bottomrule
\end{tabular*}
\footnotetext{Note: Index metadata retrieved from Bloomberg Terminal. “Curr.” denotes index currency.}
\end{table}

To account for exogenous influences on asset returns, we incorporate a range of macroeconomic indicators—including interest rates, inflation metrics, GDP growth rates, and employment figures—alongside global commodity prices, foreign exchange rates, and investor sentiment signals derived from structured news feeds, social media analytics, insider trading, and options data. These features enrich the state representation with both fundamental and behavioral dimensions of market dynamics. A full taxonomy of input features retrieved from Bloomberg Terminal is provided in Table~\ref{tab:bloomberg_features}.

The core equity universe of our study is composed of seven major stock indices, summarized in Table~\ref{tab:indices}. These indices span a diverse set of regions---including North America, Europe, and Asia---and capture both developed and emerging markets. Two universes (the SSE Composite with more than 2{,}200 stocks and the S\&P~500 with 500 constituents) meet or exceed the 500-name threshold, so the agent operates over a large cross-section on those markets. Their inclusion ensures exposure to a variety of liquidity regimes, market structures, and economic conditions.

Each index reflects distinct weighting methodologies. For example, the S\&P 500 and FTSE 100 are free-float market capitalization weighted, offering broad exposure to large-cap equities in the U.S. and U.K., respectively. The DAX and Hang Seng indices also follow cap-weighted schemes but represent concentrated baskets of domestic blue-chip firms. In contrast, the Dow Jones Industrial Average (DJIA) is price-weighted, while the NASDAQ-100 uses a modified capitalization approach to reduce concentration risk in the tech-heavy composition. The SSE Composite, representing nearly the entire Chinese stock market, updates daily and captures high-frequency volatility in an emerging market context.

Index sizes range from as few as 30 constituents (DJIA) to more than 2{,}200 (SSE Composite), and rebalance schedules vary from quarterly to irregular intervals. This diversity lets us evaluate the scalability and generalization of our approach across varying breadth, turnover dynamics, and market microstructures.

Furthermore, index-level and constituent-level data—such as returns, prices, volumes, and corporate actions—are integrated into the model’s state space and benchmarking logic. For \textasciicircum{}GSPC, \textasciicircum{}NDX, \textasciicircum{}FTSE, \textasciicircum{}GDAXI, \textasciicircum{}HSI, and \textasciicircum{}DJI, Bloomberg historical membership feeds are available and we therefore use time-varying constituent sets within each walk-forward split. For 000001.SS, historical constituent history is incomplete in our licensed feed, so we use a fixed Bloomberg constituent snapshot as of 2025-02-01 as a proxy universe. Table~\ref{tab:membership_sources} summarizes this treatment. Indices that meet or exceed the 500-name threshold (SSE Composite exceeds; S\&P~500 meets) therefore force the agent to optimize over very large cross-sections, confirming the intended scaling regime. Training and evaluating across this geographically diverse index panel exposes the method to heterogeneous liquidity regimes, sectoral compositions, market capitalizations, and rebalancing schedules.

\subsubsection{Descriptive Statistics}

To characterize the distributional properties of the underlying equity markets, we report standard summary statistics for daily log returns of each benchmark index in Table~\ref{tab:descriptive_stats}. Consistent with the walk-forward protocol, these descriptive statistics (and all downstream inferences that reference them) are computed exclusively on the 2014--2024 out-of-sample evaluation window; pre-2014 observations merely provide training context and never enter the table. For each series we list the number of observations ($N$), mean and median return, standard deviation, extrema, skewness, excess kurtosis (fourth standardized moment minus three, so that a normal distribution has excess kurtosis $=0$), and the $p$-value of a Jarque--Bera normality test. This provides a compact view of the heavy tails and volatility clustering that are typical of equity returns and motivates the use of risk-aware performance measures in our reinforcement learning framework.

\begin{table*}[!htbp]
\centering
\caption{Descriptive statistics of daily log returns for the benchmark indices over the 2014--2024 walk-forward evaluation window. ``JB $p$'' reports the $p$-value of a Jarque--Bera test for normality. All returns are expressed in the local index currency.}\label{tab:descriptive_stats}
\scriptsize
\setlength{\tabcolsep}{3pt} 
\begin{tabular}{p{1.27cm}rrrrrrrrr}
\toprule
\textbf{Index} & \textbf{$N$} & \textbf{Mean} & \textbf{Median} & \textbf{Std} & \textbf{Min} & \textbf{Max} & \textbf{Skew} & \textbf{Ex.\ Kurt.} & \textbf{JB $p$} \\
\midrule
000001.SS          & 2708 & 0.00017 & 0.00047 & 0.01298 & -0.08873 & 0.07755 & -1.008 & 8.319 & $<0.001$ \\
\textasciicircum{}DJI & 2797 & 0.00036 & 0.00066 & 0.01074 & -0.13842 & 0.10764 & -0.945 & 23.322 & $<0.001$ \\
\textasciicircum{}FTSE   & 2814 & 0.00009 & 0.00054 & 0.00967 & -0.11512 & 0.08667 & -0.870 & 13.364 & $<0.001$ \\
\textasciicircum{}GDAXI       & 2820 & 0.00031 & 0.00080 & 0.01195 & -0.13055 & 0.10414 & -0.591 & 10.293 & $<0.001$ \\
\textasciicircum{}GSPC   & 2797 & 0.00043 & 0.00067 & 0.01093 & -0.12765 & 0.08968 & -0.812 & 16.111 & $<0.001$ \\
\textasciicircum{}HSI   & 2742 & -0.00001 & 0.00026 & 0.01330 & -0.09879 & 0.08693 & -0.019 &  3.776 & $<0.001$ \\
\textasciicircum{}NDX & 2797 & 0.00065 & 0.00117 & 0.01352 & -0.13003 & 0.09597 & -0.537 & 7.361 & $<0.001$ \\
\bottomrule
\end{tabular}
\normalsize
\end{table*}

Mean daily returns are small and mostly mildly positive (\textasciicircum{}HSI is essentially flat at $-0.00001$), with markedly fat-tailed and consistently left-skewed distributions; Jarque--Bera tests reject normality at any conventional significance level even when computed strictly on the 2014--2024 out-of-sample horizon. A standard Engle ARCH LM test (not tabulated) further indicates pronounced conditional heteroscedasticity for all indices, consistent with well-known stylized facts such as volatility clustering in equity markets.

\subsection{Data Handling and Look-Ahead Bias}\label{subsec:data_handling}

To ensure that the backtests are implementable in real time and free from artificial information leakage, we align all predictors so that only information available at (or before) the portfolio-formation date is used.

\subsubsection{Company Fundamentals}
Quarterly and annual financial statement data (income statements, balance sheets, and cash-flow statements) obtained from Bloomberg Terminal and Financial Modeling Prep are lagged to reflect realistic reporting delays~\cite{bloomberg2025, fmp2025}. Concretely, we assume that each statement becomes tradable information only after a fixed lag of 60 calendar days following the period end; all fundamental features are shifted accordingly, so that values released for quarter $q$ first enter the feature set in the subsequent period.

\subsubsection{Earnings and Analyst Forecasts}
Earnings surprises and analyst forecast revisions are constructed so that only contemporaneously available estimates enter the state. To avoid look-ahead bias from backfilled analyst databases, we treat revisions as occurring at their recorded announcement dates and do not use future revisions when computing past signals. Earnings announcements that occur after the market close can only affect trading decisions from the next business day onward.

\subsubsection{Macroeconomic and Survey Data}
Macroeconomic indicators (e.g.\ interest rates, inflation measures, industrial production, and unemployment) are used at a monthly or quarterly frequency and are aligned to their official release dates. Where real-time vintages are not available, we conservatively lag macro variables by one publication period so that revisions cannot influence earlier decisions.

\subsubsection{Index Constituents and Survivorship}
Index membership, corporate actions, and free-float shares are sourced from Bloomberg. Table~\ref{tab:membership_sources} lists the exact membership treatment by index. Historical time-varying membership sets are used for \textasciicircum{}GSPC, \textasciicircum{}NDX, \textasciicircum{}FTSE, \textasciicircum{}GDAXI, \textasciicircum{}HSI, and \textasciicircum{}DJI\@. For 000001.SS, where full historical constituent history is not available in our licensed extract, we use the published constituent snapshot as of 2025-02-01 as a fixed proxy universe; this is the main survivorship-bias caveat in our setup.

\begin{table}[htbp]
\small
\caption{Index-membership sourcing and survivorship treatment used in backtests.}\label{tab:membership_sources}
\setlength{\tabcolsep}{4pt}
\begin{tabular}{p{1.7cm}p{2.6cm}p{6.9cm}}
\toprule
\textbf{Index} & \textbf{Membership mode} & \textbf{Timestamp / note} \\
\midrule
000001.SS & Static proxy & Bloomberg constituent snapshot as of 2025-02-01. \\
\textasciicircum{}GSPC & Time-varying & Bloomberg constituent history queried by trading date. \\
\textasciicircum{}NDX & Time-varying & Bloomberg constituent history queried by trading date. \\
\textasciicircum{}FTSE & Time-varying & Bloomberg constituent history queried by trading date. \\
\textasciicircum{}GDAXI & Time-varying & Bloomberg constituent history queried by trading date. \\
\textasciicircum{}HSI & Time-varying & Bloomberg constituent history queried by trading date. \\
\textasciicircum{}DJI & Time-varying & Bloomberg constituent history queried by trading date. \\
\bottomrule
\end{tabular}
\end{table}

\subsubsection{Trading Calendars and Holidays}
All time series are aligned to exchange-specific trading calendars: non-trading days are removed rather than forward-filled, and each market is processed on its own local calendar. This avoids spurious returns on holidays and ensures that cross-market features do not inadvertently incorporate future information through asynchronous closing times.

\subsubsection{Close-to-Close Trading Rule}
Our main experiments assume a daily rebalancing scheme in which information up to and including day $t$ is used to set portfolio weights for the next trading day $t{+}1$, implying execution at (or near) the close of $t{+}1$. Transaction costs use a deterministic schedule: 5 bps per side for top-decile U.S. names (\textasciicircum{}GSPC/\textasciicircum{}NDX/\textasciicircum{}DJI), 10 bps for top-decile names in the U.K., Germany, and Hong Kong SAR, China (\textasciicircum{}FTSE/\textasciicircum{}GDAXI/\textasciicircum{}HSI), 15 bps for the remaining U.S./U.K./German names, 20 bps for the remaining Hang Seng names, and 30 bps for 000001.SS constituents. Top decile is defined by trailing 60-trading-day average dollar volume ($\mathrm{ADV}_{60} = \text{price} \times \text{volume}$) within each index universe, computed with information available up to the rebalance date and refreshed monthly (first rebalance day of each month). Borrow fees are 30 bps/year (U.S.), 45 bps/year (U.K./Germany), 75 bps/year (Hong Kong SAR, China), and 120 bps/year (China), accrued linearly over holding days. The exact constants are listed in Table~\ref{tab:cost_schedule} (Appendix~\ref{app:costs}). These charges are debited on every rebalance so that the close-to-close convention reflects adverse drift associated with delayed fills.

\subsection{Baselines}

We benchmark AlphaZeroBeta against three portfolio strategies, each based on a different optimization methodology.

\subsubsection{Index Buy-and-Hold (Index B\&H)}

The passive benchmark maintains the initial allocation unchanged over the entire evaluation period:
\begin{equation}
   w_i(t) = w_i(0), \quad \forall t,
\end{equation}
where $w_i(t)$ denotes the portfolio weight of asset $i$ at time $t$. The benchmark provides a baseline for market-tracking performance~\cite{perold_sharpe_1988}.

\subsubsection{Maximum Sharpe Ratio Portfolio (MxSharpe)}

The procedure is formulated to maximize the portfolio’s Sharpe ratio subject to linear constraints:
\begin{equation}
   \max_{w} \quad \frac{\mathbb{E}[R_p - R_f]}{\sigma_p} \quad \text{s.t.} \quad \sum_{i} w_i = 1, \quad -1 \leq w_i \leq 1,
\end{equation}
where $R_p = w^\top R$ is the portfolio return and $\sigma_p = \sqrt{w^\top \Sigma w}$ denotes portfolio volatility. The mean--variance formulation follows~\cite{markowitz1952}; the resulting program is solved via Sequential Least Squares Programming (SLSQP), which handles the nonlinear Sharpe ratio objective subject to the box and full-investment constraints.

\subsubsection{Minimal Correlation Portfolio (Decorr)}

The strategy minimizes the absolute correlation with the benchmark index to drive the portfolio toward $\rho_p \approx 0$ rather than $\rho_p \to -1$:
\begin{equation}
   \min_{w} \quad \lvert \rho_p \rvert = \left\lvert \frac{\text{Cov}(R_p, R_b)}{\sigma_p \sigma_b} \right\rvert \quad \text{s.t.} \quad \sum_{i} w_i = 1, \quad -1 \leq w_i \leq 1,
\end{equation}
where $R_b$ is the benchmark return. Because $|\rho_p|$ is non-smooth at zero, in practice we solve the equivalent smooth program $\min_w \rho_p^2$ with SLSQP, which yields the same optimizer and ensures feasible allocations with bounded leverage and full investment.

These three benchmarks span the most common convex constructions used in practice---passive, Sharpe-optimal, and decorrelated---and therefore provide an apples-to-apples reference for AlphaZeroBeta's performance in both absolute and relative terms.

The baselines and AlphaZeroBeta differ in one important respect: their constraint sets. All convex baselines use the same box constraints ($-1 \leq w_i \leq 1$) together with a full-investment condition $\sum_i w_i = 1$, so their net exposure is always long one dollar. They therefore remain sensitive to market beta unless neutrality emerges implicitly from the optimizer (only the Decorr formulation indirectly pushes toward low correlation). For AlphaZeroBeta we impose the identical per-asset bounds, project the centered policy outputs onto the $\ell_1$ ball of radius one, and therefore enforce both $\sum_i w_i = 0$ and $\sum_i |w_i| \le 1$ at every rebalance (Listing~\ref{lst:alphazerobeta-env}). The correlation penalty in Eq.~\eqref{eq:reward_full} still matters—it discourages residual beta after transaction costs—but neutrality is hard-coded rather than emergent.

This fundamental difference in constraints means the comparison favors baselines in upward-trending markets (they retain positive beta) while AlphaZeroBeta must generate performance solely from relative-value signals. The comparison is leverage-neutral in terms of gross exposure yet differs in net exposure: baselines are net-long by design, whereas AlphaZeroBeta is dollar-neutral by construction. This limitation should be considered when interpreting performance comparisons in Section~\ref{sec:results}.

To verify that our advantages are not solely attributable to using RL, we also report an ablation in Table~\ref{tab:rl_baseline} (Appendix~\ref{app:ablations}) that removes the neutrality penalty from the policy reward while leaving the architecture and training schedule unchanged. The return-maximizing RL baseline outperforms the passive index in every market but falls short of AlphaZeroBeta on Sharpe ratio and exhibits materially higher market correlations and deeper drawdowns, confirming that AlphaZeroBeta's neutrality-aware formulation, rather than RL alone, drives the improvements highlighted in Section~\ref{sec:results}.

\subsection{Sample Efficiency and Overfitting Mitigation}

Our framework improves sample efficiency and mitigates overfitting by exploiting the structure of financial markets rather than relying on synthetic data. We employ three techniques:

\begin{enumerate}
    \item At each time step, the state representation incorporates both index-level variables (e.g.\ returns, prices, realized volatility) and constituent-level features for every member of the traded universe (2{,}200 Shanghai stocks, 500 S\&P constituents, 100 FTSE names, etc.). This expansion increases the effective sample size by treating each asset as a distinct learning signal, thereby enhancing statistical power and reducing the likelihood of overfitting to idiosyncratic noise.

    \item We implement a rolling-window (moving-window) evaluation scheme with overlapping training windows across folds and disjoint validation/test slices~\cite{bailey2017}. For each split, the agent is trained on a fixed-length historical segment, validated on a held-out portion of that segment for hyperparameter tuning and early stopping, and then tested strictly on the subsequent out-of-sample period. This design exposes the agent to heterogeneous market regimes (bull, bear, sideways) and ensures that all performance metrics reflect unseen data. Profit-and-loss plots and all tabulated results are computed exclusively on out-of-sample periods, excluding the training and validation segments.

    \item Standard machine learning techniques, including dropout~\cite{srivastava2014}, weight decay~\cite{krogh1992}, and early stopping~\cite{prechelt1998}, are applied during training. These methods constrain model complexity and prevent memorization of training noise, further improving generalization to unseen data.
\end{enumerate}

\medskip
\noindent We rely solely on historical market data for both training and evaluation. No synthetic data generators or artificial perturbations are employed, ensuring that all observed behaviors and performance metrics are grounded in real market dynamics.

\section{Compute and Evaluation Protocol}\label{sec:compute}

This section consolidates the experimental infrastructure used in the empirical study: hardware and runtime requirements, evaluation metrics, and the walk-forward validation scheme (Section~\ref{sec:walk_forward_validation}). All performance metrics (PnL charts and tables) are evaluated strictly on the out-of-sample test windows defined by the walk-forward protocol.

\subsection{Hardware}
Training and validation were executed on a server-class GPU machine (Linux, CUDA-enabled), comparable to an NVIDIA DGX A100 system (8$\times$A100 GPUs, server-grade CPUs, high-bandwidth NVLink interconnect) or, for single-GPU ablations, a cloud instance with one datacenter GPU (e.g.\ AWS \texttt{p3.2xlarge}: 1$\times$V100 16\,GB, 8 vCPUs, 61\,GiB RAM). These platforms are representative of hardware commonly used in contemporary DRL-for-finance workloads and documented in vendor benchmark guides.

\subsection{Accounting of Compute}

\subsubsection{AlphaZeroBeta}

The agent is trained from scratch at each walk-forward fold and evaluated on the subsequent out-of-sample segment. The minimum hardware requirement is one datacenter-grade GPU of the V100/A100 class with at least 16\,GB of VRAM and eight CPU cores; additional GPUs enable fold- and seed-level parallelism.

On the reference single-GPU setup (AWS \texttt{p3.2xlarge}, 1$\times$V100 16\,GB), one complete train/validation cycle takes $2.6 \pm 0.3$ hours for the smaller universes (\textasciicircum{}DJI/\textasciicircum{}GDAXI/\textasciicircum{}HSI/\textasciicircum{}FTSE/\textasciicircum{}NDX, ranging from 30 to 100 names) and the 500-name \textasciicircum{}GSPC, and $4.8 \pm 0.5$ hours for the 2{,}200-name SSE Composite. On an 8$\times$A100 DGX-class node, running one fold per GPU reduces the per-fold wall-clock to $1.1$ and $2.0$ hours, respectively. Aggregating across the $K=22$ rolling splits (Jan 2014--Dec 2024), a single-seed sweep over all 22 folds of one index takes 57--106\,h on a single V100 (about 2.4--4.4 wall-clock days, depending on universe size) or 3--6\,h on the 8$\times$A100 node when the 22 folds are spread one-per-GPU across the eight A100s.

For exploratory single-seed studies we budget approximately one week of training per index on a single V100-equivalent GPU, so a single-seed pass over all seven indices takes on the order of two months end-to-end. The full evaluation reported in this paper uses nine independent seeds per fold and is therefore run on the 8$\times$A100 DGX-class node, where parallelized folds and seeds complete the entire seven-index, nine-seed sweep in approximately 12 days of wall-clock time.

\subsubsection{Benchmarks (Index B\&H, MxSharpe, Decorr)}
All three baselines are CPU-bound and entail low computational overhead:
\begin{enumerate}
  \item Index B\&H is a trivial vectorized backtest in $O(TN)$ complexity.
  \item MxSharpe and Decorr are smooth nonlinear programs per rebalance (MxSharpe maximizes a ratio of mean to volatility; Decorr minimizes squared benchmark correlation), each with linear box and budget constraints and solved with SLSQP; for $N \sim {\mathcal O}(10^2\!-\!10^3)$ assets and typical rebalance grids, per-window solves complete in seconds, so full walk-forward backtests over multiple indices comfortably fit within $\leq$1 day on a modern multi-core CPU\@.
\end{enumerate}

\subsection{Evaluation Metrics}

To assess portfolio performance we report six metrics, grouped by role. Three are used for tabular comparison across methods: the Sharpe ratio (risk-adjusted return), the Pearson correlation with the benchmark (neutrality diagnostic), and the maximum drawdown (downside-risk measure)~\cite{sharpe1966, perold_sharpe_1988, lo2002}. Three are used for visual diagnostics in the PnL charts: cumulative return, net exposure (long minus short), and gross exposure (long plus short).

\subsubsection{Sharpe Ratio}
The Sharpe ratio $S$ measures the excess return per unit of volatility:
\begin{equation}
  S = \frac{\mathbb{E}[R_p - R_f]}{\sigma_p},
\end{equation}
where $R_p$ is the portfolio return, $R_f$ is the risk-free rate, and $\sigma_p$ is the standard deviation of $R_p$. A higher Sharpe ratio indicates superior risk-adjusted performance~\cite{sharpe1994}.
In all reported results we use daily excess returns over the effective federal funds rate (lagged to avoid look-ahead), and we annualize Sharpe by multiplying the daily estimate by $\sqrt{252}$.

\subsubsection{Correlation with Benchmark}
The Pearson correlation $\rho$ between portfolio returns $R_p$ and benchmark returns $R_b$ is defined as:
\begin{equation}
  \rho = \frac{\text{Cov}(R_p, R_b)}{\sigma_p \sigma_b},
\end{equation}
where $\sigma_p$ and $\sigma_b$ are the respective standard deviations. This metric evaluates the degree of market neutrality; lower values imply greater independence from the index~\cite{patton2009}.

\subsubsection{Maximum Drawdown}
Maximum drawdown (MDD) quantifies the largest historical peak-to-trough loss during the investment horizon:
\begin{equation}
  \text{MDD} = \min_{t \in [0, T]} \left( \frac{V_t - V_{\max}}{V_{\max}} \right),
\end{equation}
where $V_t$ is the portfolio value at time $t$ and $V_{\max} = \max_{\tau \leq t} V_\tau$ is the running maximum. MDD provides a measure of capital risk under worst-case historical scenarios~\cite{chekhlov2005}.

\subsubsection{Cumulative Return}
Cumulative return measures the total profit or loss accrued over time without accounting for risk. It is computed as the cumulative sum of portfolio-level profit and loss:
\begin{equation}
  \mathcal{C}_t = \sum_{s=1}^{t} \Delta P_s,
\end{equation}
where $\Delta P_s$ denotes the change in portfolio value at time $s$, computed as the inner product of asset price changes and lagged position sizes. The statistic $\mathcal{C}_t$ serves as a raw indicator of strategy performance over the investment horizon~\cite{campbell1997}.

\subsubsection{Market Value (Net Exposure)}
Net exposure reflects the portfolio’s directional market stance. At time $t$, it is defined as:
\begin{equation}
  \mathcal{E}^{\text{net}}_t = \sum_{i=1}^{N} p_{i,t} \cdot q_{i,t},
\end{equation}
where $p_{i,t}$ is the price of asset $i$ and $q_{i,t}$ is the signed quantity held. Positive values indicate long exposure; negative values indicate short positioning~\cite{grinold2000}.

\subsubsection{Gross Value (Gross Exposure)}
Gross exposure captures the total capital deployed, regardless of position direction:
\begin{equation}
  \mathcal{E}^{\text{gross}}_t = \sum_{i=1}^{N} p_{i,t} \cdot \lvert q_{i,t} \rvert.
\end{equation}
This metric reflects leverage and portfolio scale, and is critical for risk and margin evaluation in long--short and market-neutral strategies~\cite{asness2012}.

\medskip
\noindent Note: All metrics are computed assuming an initial capital base of 1 unit in the benchmark index's currency (e.g.\ \$1), with method-specific exposure constraints applied at each rebalancing date and no external cash flows.

\subsection{Walk-Forward Validation}\label{sec:walk_forward_validation}

The walk-forward validation protocol introduced in Section~\ref{sec:experiment} ensures robust generalization and prevents temporal leakage. During the out-of-sample phase we simulate daily rebalancing with the deterministic transaction-cost and borrow-fee schedule reported in Table~\ref{tab:cost_schedule}, always using information available up to each decision point.

Each of the $K=22$ walk-forward folds (Section~\ref{sec:experiment}) consists of a 36-month training window, a 6-month validation window used for early stopping and model selection, and a non-overlapping 6-month test window over which performance is reported. The block advances in 6-month steps, so consecutive training windows overlap by 30 months while the validation and test slices advance disjointly. The 22 test windows tile January 2014 through December 2024 without overlap, providing a synchronized evaluation horizon across markets with heterogeneous holiday calendars; pre-2014 data are reserved exclusively for warm-up of rolling estimators (Section~\ref{sec:experiment}) and never enter the reported metrics.

The cumulative profit-and-loss (PnL) chart (Figure~\ref{fig:pnl_main}) and the performance metrics in Table~\ref{tab:strategy_comparison}---annualized Sharpe ratio, maximum drawdown, and correlation to the benchmark index---are computed over the 22 out-of-sample test windows. Unless stated otherwise, we report mean $\pm$ standard deviation across the Cartesian product of the $K=22$ folds and nine independent RL initializations per fold (different random seeds for the policy and value networks under identical hyperparameters), so the dispersion columns capture both temporal variability and sensitivity to agent restarts. This design emphasizes generalization across time and mitigates overfitting to specific market regimes~\cite{harvey2015}.

\section{Results}\label{sec:results}

Table~\ref{tab:strategy_comparison} summarizes the comparative performance of AlphaZeroBeta across the seven equity benchmarks described in Table~\ref{tab:indices}, using the 2014--2024 out-of-sample segments generated by the walk-forward procedure. AlphaZeroBeta is evaluated against three reference strategies:

\begin{enumerate}
  \item \textbf{Index B\&H}: A passive buy-and-hold strategy on the benchmark index, representing the baseline market exposure.
  \item \textbf{Decorr}: A constrained optimization strategy that minimizes correlation with the benchmark index.
  \item \textbf{MxSharpe}: A return-risk optimized strategy that maximizes the Sharpe ratio without imposing market neutrality constraints.
\end{enumerate}

\begin{table}[htbp]
\caption{Performance comparison across methods. For each index, we highlight in bold: the best Sharpe ratio (highest), best max drawdown (closest to zero), and lowest absolute correlation.}\label{tab:strategy_comparison}
\begin{tabular*}{\textwidth}{@{\extracolsep\fill}llccc}
\toprule
\textbf{Index} & \textbf{Method} & \textbf{Sharpe ratio} & \textbf{Max drawdown} & \textbf{Correlation} \\
\midrule
\multirow{4}{*}{000001.SS}
 & \textit{AlphaZeroBeta}     & \textbf{1.63 ± 0.38} & -0.34 ± 0.11 & \textbf{-0.02 ± 0.05} \\
 & Index B\&H        & 0.30 ± 0.00 & -0.71 ± 0.00 & 1.00 ± 0.00 \\
 & Decorr            & -0.24 ± 0.20 & \textbf{-0.24 ± 0.08} & -0.05 ± 0.21 \\
 & MxSharpe          & 0.80 ± 0.26 & -0.63 ± 0.36 & 0.89 ± 0.09 \\
\midrule
\multirow{4}{*}{\textasciicircum{}FTSE}
 & \textit{AlphaZeroBeta}     & \textbf{0.94 ± 0.19} & -0.28 ± 0.09 & \textbf{-0.02 ± 0.08} \\
 & Index B\&H        & 0.23 ± 0.00 & -0.43 ± 0.00 & 1.00 ± 0.00 \\
 & Decorr            & -0.07 ± 0.18 & -0.44 ± 0.07 & -0.02 ± 0.17 \\
 & MxSharpe          & 0.37 ± 0.16 & \textbf{-0.22 ± 0.20} & 0.73 ± 0.21 \\
\midrule
\multirow{4}{*}{\textasciicircum{}GDAXI}
 & \textit{AlphaZeroBeta}     & \textbf{0.86 ± 0.23} & \textbf{-0.16 ± 0.11} & \textbf{0.05 ± 0.04} \\
 & Index B\&H        & 0.51 ± 0.00 & -0.57 ± 0.00 & 1.00 ± 0.00 \\
 & Decorr            & -0.16 ± 0.17 & -0.49 ± 0.12 & -0.10 ± 0.23 \\
 & MxSharpe          & 0.11 ± 0.25 & -0.28 ± 0.21 & 0.44 ± 0.35 \\
\midrule
\multirow{4}{*}{\textasciicircum{}GSPC}
 & \textit{AlphaZeroBeta}     & \textbf{1.61 ± 0.48} & -0.26 ± 0.15 & 0.15 ± 0.09 \\
 & Index B\&H        & 0.72 ± 0.00 & -0.67 ± 0.00 & 1.00 ± 0.00 \\
 & Decorr            & -0.27 ± 0.17 & -0.88 ± 0.14 & \textbf{-0.11 ± 0.19} \\
 & MxSharpe          & 0.91 ± 0.44 & \textbf{-0.09 ± 0.31} & 0.92 ± 0.07 \\
\midrule
\multirow{4}{*}{\textasciicircum{}HSI}
 & \textit{AlphaZeroBeta}     & \textbf{1.04 ± 0.33} & -0.21 ± 0.22 & \textbf{0.01 ± 0.04} \\
 & Index B\&H        & 0.10 ± 0.00 & -0.67 ± 0.00 & 1.00 ± 0.00 \\
 & Decorr            & -0.19 ± 0.18 & -0.63 ± 0.16 & -0.07 ± 0.18 \\
 & MxSharpe          & 0.62 ± 0.31 & \textbf{-0.18 ± 0.27} & 0.81 ± 0.13 \\
\midrule
\multirow{4}{*}{\textasciicircum{}NDX}
 & \textit{AlphaZeroBeta}     & \textbf{1.48 ± 0.41} & -0.32 ± 0.10 & \textbf{0.07 ± 0.06} \\
 & Index B\&H        & 0.90 ± 0.00 & -0.75 ± 0.00 & 1.00 ± 0.00 \\
 & Decorr            & -0.12 ± 0.15 & -0.53 ± 0.13 & -0.08 ± 0.22 \\
 & MxSharpe          & 0.85 ± 0.37 & \textbf{-0.19 ± 0.26} & 0.86 ± 0.11 \\
\midrule
\multirow{4}{*}{\textasciicircum{}DJI}
 & \textit{AlphaZeroBeta}     & \textbf{1.20 ± 0.28} & -0.27 ± 0.14 & \textbf{0.03 ± 0.07} \\
 & Index B\&H        & 0.58 ± 0.00 & -0.53 ± 0.00 & 1.00 ± 0.00 \\
 & Decorr            & -0.18 ± 0.19 & -0.46 ± 0.11 & -0.06 ± 0.15 \\
 & MxSharpe          & 0.76 ± 0.29 & \textbf{-0.22 ± 0.24} & 0.79 ± 0.13 \\
\bottomrule
\end{tabular*}
\footnotetext{Note: Results are expressed as mean ± standard deviation. For AlphaZeroBeta, dispersion is computed across the Cartesian product of 22 walk-forward folds and nine independent RL agent initializations per fold (198 samples). Index B\&H reports a single buy-and-hold path computed over the full 2014--2024 horizon and therefore has zero dispersion. Decorr and MxSharpe are deterministic per fold; their reported dispersion captures fold-to-fold variation across the 22 walk-forward windows (no agent-restart variation, because the optimizers are deterministic). Bold entries highlight the best Sharpe ratio (highest), best max drawdown (closest to zero), and lowest absolute correlation for each index.}
\end{table}

On average AlphaZeroBeta delivers a Sharpe ratio of \(1.25\) (cross-market standard deviation \(0.30\)), computed as the simple mean and standard deviation of the per-market entries in Table~\ref{tab:strategy_comparison}; the best convex baseline in each market averages \(0.70\) (standard deviation \(0.19\)). AlphaZeroBeta also maintains correlations within \(\pm 0.15\) of zero in every region.

Our method achieves the highest Sharpe ratio in all markets, suggesting superior risk-adjusted performance. Drawdown is competitive but not uniformly best: AlphaZeroBeta has the smallest (most positive) maximum drawdown only on \textasciicircum{}GDAXI, while MxSharpe or Decorr record less-severe drawdowns in the other markets (numbers are reported as negative fractions of peak equity). In highly liquid indices such as the S\&P 500 (\textasciicircum{}GSPC) and Nasdaq-100 (\textasciicircum{}NDX), AlphaZeroBeta achieves Sharpe ratios of 1.61 and 1.48, respectively—substantially outperforming benchmarks.

The low or near-zero correlation with the corresponding index B\&H further confirms that our strategy learns non-trivial return structures, providing strong orthogonality to passive exposure. In some cases, such as 000001.SS and \textasciicircum{}FTSE, the realized correlation is slightly negative or indistinguishable from zero, indicating that the soft correlation penalty in the reward suffices to neutralize residual beta without requiring a hard zero-correlation constraint.

The occasional underperformance of the ex-ante maximum-Sharpe (MxSharpe) portfolio arises from the sensitivity of mean--variance optimization to estimation noise and non-stationary return moments. As a result, the theoretically optimal in-sample weights degrade out-of-sample, particularly in indices with unstable covariance structure or high concentration such as \textasciicircum{}GDAXI and \textasciicircum{}NDX\@.

The convex baselines remain fully invested and net long, whereas AlphaZeroBeta is dollar-neutral by construction (zero-sum weights with an \(\ell_1\)-projection). The comparison therefore favors the baselines whenever the underlying index trends upward because they retain positive beta, while AlphaZeroBeta must generate performance from relative-value signals alone. Decorr partially mitigates this imbalance by minimizing realized correlation, yet even that formulation lacks an explicit zero-sum constraint. The fact that AlphaZeroBeta still outperforms highlights that its hard neutrality constraint plus the neutrality-aware reward prevent beta drift without sacrificing Sharpe ratio.

Transaction costs and borrow fees follow the deterministic schedule reported in Table~\ref{tab:cost_schedule}. Under this schedule the realized absolute turnover $\sum_i |\Delta w_i|$ for AlphaZeroBeta averages $0.56 \pm 0.11$ per rebalance across markets (95th percentile $0.92$), translating into an all-in implementation cost of roughly 8--12 bps per day in the most active windows. These magnitudes indicate that the reward penalty produces realistic trading intensity while leaving ample headroom for net alpha after frictions.
The identical slippage/borrow model is applied to every strategy—Index B\&H, Decorr, MxSharpe, and AlphaZeroBeta—with costs debited on each rebalance (buy-and-hold only pays when index membership changes). All four strategies rebalance at the same daily frequency, so they incur comparable per-rebalance costs under this model: the mean turnover is $\approx 0.50$ for Decorr and $\approx 0.47$ for MxSharpe across markets (not tabulated separately), against $0.56$ for AlphaZeroBeta noted above. All reported Sharpe ratios and drawdowns are net of these cost assumptions.

To isolate the effect of AlphaZeroBeta's neutrality-aware reward design, Appendix~\ref{app:ablations} benchmarks the full agent against an otherwise identical RL policy that removes the correlation penalty and simply maximizes risk-adjusted return. The ablation achieves Sharpe ratios above the passive index in every market but exhibits materially higher drawdowns and market correlations, underscoring the importance of explicit neutrality regularization.

\subsection{Stability and Generalization}

The per-market dispersions in Table~\ref{tab:strategy_comparison} reflect AlphaZeroBeta's stability across walk-forward folds and agent restarts; the modest cross-market spread (Sharpe std $0.30$) and the consistently near-zero correlations together suggest that the policy generalizes across heterogeneous regimes---from developed markets such as \textasciicircum{}GDAXI to emerging markets such as 000001.SS---rather than fitting to a particular geography.

\begin{figure}[htbp]
\centering
\begin{tikzpicture}
\begin{groupplot}[
  group style={group size=2 by 1, horizontal sep=14mm},
  finance base,
  height=6.0cm,
  width=0.46\linewidth,
  xmin=0, xmax=35000,
  scaled x ticks=false,
  xtick={0,10000,20000,30000},
  xticklabels={$0$,{$10k$},{$20k$},{$30k$}},
  xlabel={Optimization step},
  tick align=outside,
]
\nextgroupplot[
  ylabel={Policy gradient loss},
  ymin=-0.062, ymax=-0.018,
  ytick={-0.06,-0.05,-0.04,-0.03,-0.02},
  scaled y ticks=false,
  yticklabel={$\pgfmathprintnumber[fixed,precision=2]{\tick}$},
]
\foreach \col in {ppo_1,ppo_2,ppo_3,ppo_4,ppo_5,ppo_6,ppo_7,ppo_8,ppo_9}{
  \addplot[cBlue!30, line width=0.5pt, forget plot]
    table[col sep=comma, x=step, y=\col]{figures/data/training_loss.csv};
}
\addplot[black, line width=0.9pt]
  table[col sep=comma, x=step, y=mean]{figures/data/training_loss.csv};
\nextgroupplot[
  ylabel={Explained variance $R^2$},
  ymin=-23, ymax=1.15,
  ytick={-20,-10,0,1},
  scaled y ticks=false,
]
\foreach \col in {ppo_1,ppo_2,ppo_3,ppo_4,ppo_5,ppo_6,ppo_7,ppo_8,ppo_9}{
  \addplot[cBlue!30, line width=0.5pt, forget plot]
    table[col sep=comma, x=step, y=\col]{figures/data/training_ev.csv};
}
\addplot[black, line width=0.9pt]
  table[col sep=comma, x=step, y=mean]{figures/data/training_ev.csv};
\draw[cGrid, dashed, line width=0.5pt]
  (axis cs:0,0) -- (axis cs:35000,0);
\draw[cGrid, dashed, line width=0.5pt]
  (axis cs:0,1) -- (axis cs:35000,1);
\node[anchor=north east, font=\scriptsize, text=cGray]
  at (axis cs:34500,0.78) {$R^2 \approx 0.97$};
\end{groupplot}
\end{tikzpicture}
\caption{Training diagnostics across nine independent PPO runs on the S\&P~500 reference market: (left) policy gradient loss convergence over the optimization horizon; (right) explained variance of the value function over training steps. Thin blue lines: per-seed trajectories; thick black line: cross-seed mean.}\label{fig:training_stability}
\end{figure}
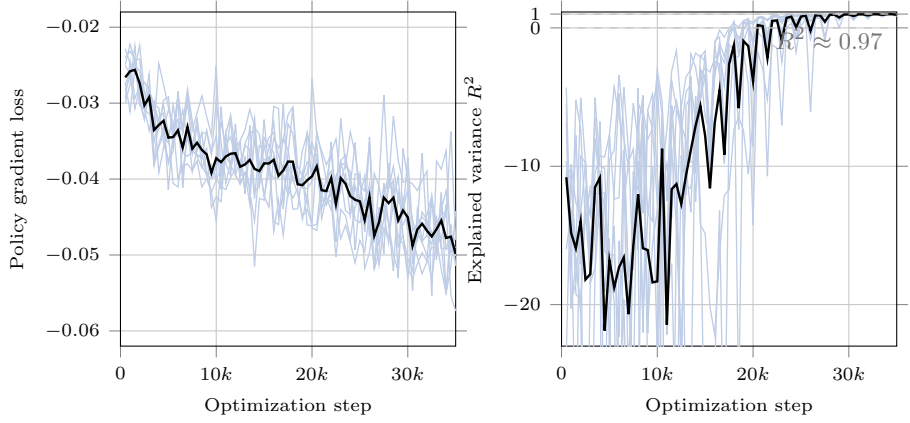

Figure~\ref{fig:training_stability} presents training diagnostics for AlphaZeroBeta across nine independently seeded PPO runs. The left panel shows the policy gradient loss stabilizing as training progresses, with convergence behavior consistent across all seeds. The variation in early training smooths out after 20k steps, suggesting reliable optimization behavior across random initializations.

The right panel reports explained variance, which measures how well the critic predicts realized returns. Most runs converge above $0.97$ (cross-seed mean $\approx 0.95$), with a minority settling near $0.85$, indicating that the critic accurately tracks the return-to-go targets that drive the policy gradient's advantage estimates. The combination of stable losses and high $R^2$ across seeds is consistent with convergence to a neighborhood of a stationary point of the joint objective; the dispersion across seeds narrows after roughly $20{,}000$ steps.

\subsection{Cumulative Performance Illustration}

To present a clear visualization of cumulative gains and drawdown dynamics, Figure~\ref{fig:pnl_main} shows the cumulative PnL curve of AlphaZeroBeta on one representative index—\textasciicircum{}GSPC (S\&P 500)—over the full backtest period.

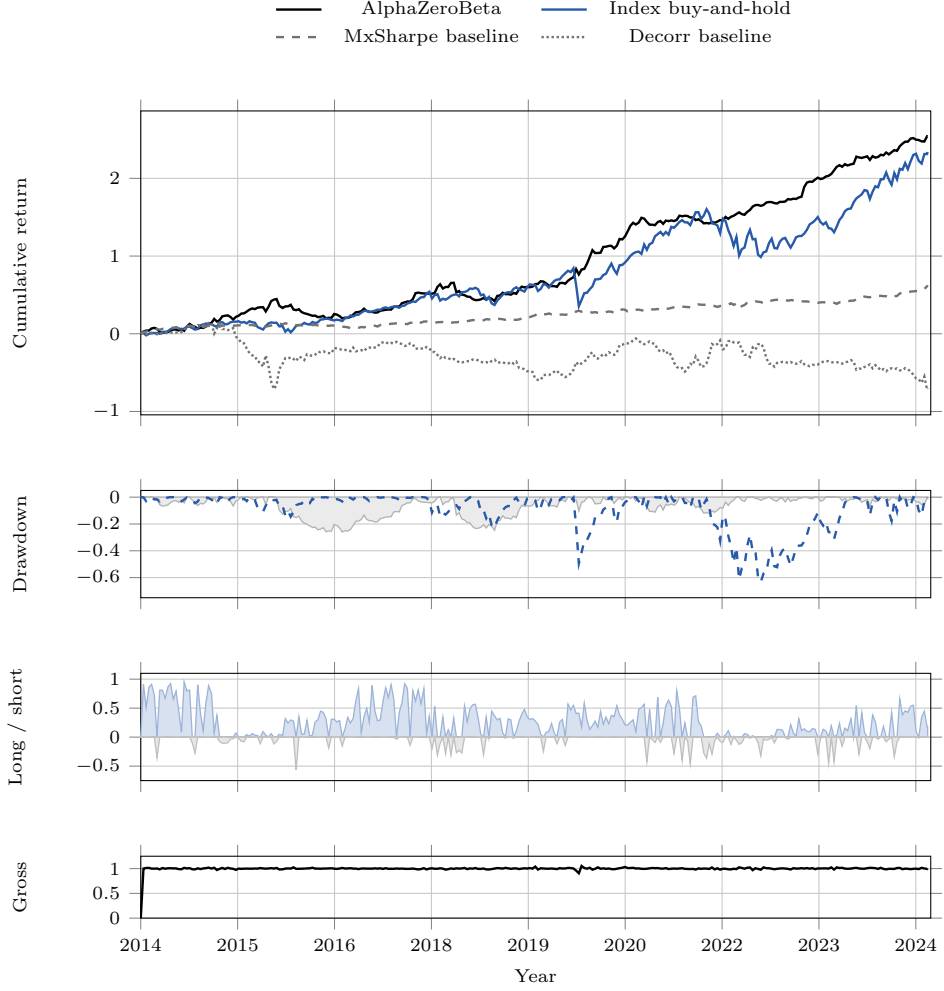
\pnlchart{figures/data/pnl_gspc.csv}{figures/data/dd_gspc.csv}{figures/data/exp_gspc.csv}{Representative AlphaZeroBeta performance on the S\&P~500 (\textasciicircum{}GSPC). Top: cumulative return for AlphaZeroBeta vs.\ index buy-and-hold, MxSharpe, and Decorr baselines. Below: drawdown, long/short exposure, and gross exposure trajectories. Returns are net of transaction costs and reflect daily rebalancing under walk-forward retraining.}{fig:pnl_main}

The strategy compounds steadily, avoids regime-specific collapses, and maintains stable return growth even through volatile periods, consistent with the per-market Sharpe ratios reported in Table~\ref{tab:strategy_comparison}.

\subsection{Full Market Performance Breakdown}

While Figure~\ref{fig:pnl_main} provides a visual overview for a single market, the full set of cumulative PnL trajectories across all benchmark indices is included in Appendix~\ref{app:charts}. These show that the agent generalizes across diverse geographies and index weighting schemes.

\subsection{Risk Factor Exposures and Attribution}\label{subsec:factor_exposures}

To assess whether AlphaZeroBeta's returns can be explained by standard risk premia, we run a factor attribution using a consistent panel of risk factors across markets. We start from the market, size, value, and profitability legs of the Fama--French five-factor model~\cite{fama2015}. The investment factor (CMA) is excluded because it is nearly collinear with the proprietary quality-minus-junk spread used in our universes, which produced unstable estimates in preliminary tests; instead we retain that quality-minus-junk proxy (QUAL) to capture cross-sectional quality tilts. On top of these factors we add the Carhart momentum factor (MOM)~\cite{carhart1997}, building on the momentum anomaly documented by \citet{jegadeesh1993}, and a short-term reversal factor (REV) following \citet{jegadeesh1990}. This combination captures trend-following, mean-reversion, and quality dynamics without duplicating exposures.

\subsubsection{Model Specification}

For each market \( p \) with available factor data (six equity markets in our study: 000001.SS, \textasciicircum{}FTSE, \textasciicircum{}GDAXI, \textasciicircum{}GSPC, \textasciicircum{}HSI, and \textasciicircum{}NDX), we regress daily excess portfolio returns on this extended set. We exclude \textasciicircum{}DJI from this attribution because, with only 30 large-cap constituents, local SMB/HML/RMW/QUAL spread construction is economically thin and statistically unstable, so coefficient estimates are not reliable. For 000001.SS, factor legs are computed on the same 2025-02-01 proxy universe described in Table~\ref{tab:membership_sources}; the resulting loadings therefore inherit the same survivorship caveat.
\begin{equation}
\begin{aligned}
r_{p,t}^{e} = \alpha_p
& + \beta_{m,p}\,r_{\mathrm{MKT},t} + \beta_{s,p}\,\mathrm{SMB}_t + \beta_{v,p}\,\mathrm{HML}_t \\
& + \beta_{q,p}\,\mathrm{RMW}_t + \beta_{\mathrm{mom},p}\,\mathrm{MOM}_t + \beta_{\mathrm{rev},p}\,\mathrm{REV}_t \\
& + \beta_{\mathrm{qual},p}\,\mathrm{QUAL}_t + \varepsilon_{p,t},
\end{aligned}
\label{eq:factor_model_extended}
\end{equation}
where the MKT/SMB/HML/RMW factor definitions follow \citet{fama1993,fama2015}, the momentum factor follows \citet{carhart1997}, and the short-term reversal follows \citet{jegadeesh1990}.
Newey--West heteroskedasticity- and autocorrelation-robust \( t \)-statistics accompany all coefficients.

Factor series are aligned carefully to avoid inadvertent lead/lag contamination. U.S. factors (Ken French library data) are merged at New York close and used contemporaneously for \textasciicircum{}GSPC and \textasciicircum{}NDX\@. For European indices (\textasciicircum{}FTSE, \textasciicircum{}GDAXI) and Asian indices (000001.SS, \textasciicircum{}HSI), where no equivalent public factor library is available, we construct MKT/SMB/HML/RMW directly from local index constituents using the standard Fama--French sort breakpoints (MKT as the cap-weighted local index over the local risk-free rate; SMB and HML from 2$\times$3 size/value sorts; RMW from operating-profitability sorts), reindex the resulting series to the local exchange calendar, and delay the returns by one trading day whenever the factor's publication timestamp occurs after the local market close. Momentum and reversal factors are constructed analogously from the same local universes and follow the same timing rule. This alignment guarantees that exposure estimates reflect implementable lead/lag conventions rather than asynchronous time stamps.

\subsubsection{Regression Results}

Table~\ref{tab:factor_results_tstats} reports the resulting coefficients for the six markets with usable factor feeds. To keep naming consistent with Table~\ref{tab:strategy_comparison}, we report Bloomberg index codes in the first column and retain the same market order (with \textasciicircum{}DJI omitted for the reason above). Market betas are statistically indistinguishable from zero across all regions, confirming that AlphaZeroBeta maintains dollar-neutral exposures. Momentum loadings are positive and significant in every market, while reversal coefficients are negative, indicating that the policy favors sustained trends over short-term noise. Quality loadings are close to zero overall but mildly negative in the NASDAQ portfolio, consistent with temporary tilts toward high-growth, lower-profitability stocks. Intercepts remain positive and significant at conventional levels, reflecting benchmark-relative abnormal returns rather than universal CAPM alpha. Asterisks in Table~\ref{tab:factor_results_tstats} follow the standard convention (* $p<0.10$, ** $p<0.05$, *** $p<0.01$).

\begin{table*}[!htbp]
\centering
\small
\setlength{\tabcolsep}{3pt}
\caption{Extended factor regression results with Newey--West \(t\)-statistics in parentheses. Specification corresponds to Eq.~\eqref{eq:factor_model_extended}. QUAL denotes our profitability-based quality-minus-junk factor constructed from ROE, gross profitability, and leverage filters following \citet{asness2019}; the Fama--French CMA factor is omitted because of near collinearity with QUAL on the universes studied.}\label{tab:factor_results_tstats}
\begin{tabular}{@{}lrrrrrrrr@{}}
\toprule
\textbf{Market} & $\alpha$ & MKT & SMB & HML & RMW & MOM & REV & QUAL \\
\midrule
000001.SS & 0.00047*** & -0.002 & 0.032 & 0.009 & 0.005 & 0.114*** & -0.029* & -0.013 \\
          & (3.15) & (-0.14) & (1.56) & (0.48) & (0.44) & (3.55) & (-1.76) & (-0.99) \\
\textasciicircum{}FTSE & 0.00029** & -0.004 & -0.017 & 0.011 & 0.014 & 0.091*** & -0.027* & -0.009 \\
                      & (2.12) & (-0.22) & (-0.88) & (0.55) & (1.02) & (3.74) & (-1.76) & (-0.77) \\
\textasciicircum{}GDAXI & 0.00033*** & 0.006 & -0.021 & 0.019 & 0.004 & 0.105*** & -0.031** & -0.022 \\
                       & (2.76) & (0.42) & (-1.02) & (1.05) & (0.41) & (4.09) & (-2.24) & (-1.19) \\
\textasciicircum{}GSPC & 0.00042*** & 0.011 & 0.028 & -0.014 & 0.006 & 0.137*** & -0.042** & -0.011 \\
                      & (3.01) & (0.62) & (1.44) & (-0.88) & (0.55) & (4.92) & (-2.11) & (-1.08) \\
\textasciicircum{}HSI & 0.00031** & 0.009 & 0.013 & 0.018 & 0.011 & 0.128*** & -0.036** & -0.016 \\
                     & (2.09) & (0.55) & (0.72) & (0.99) & (0.88) & (3.88) & (-2.01) & (-1.06) \\
\textasciicircum{}NDX & 0.00055*** & 0.019 & 0.044* & -0.022 & -0.031* & 0.162*** & -0.038* & -0.067** \\
                     & (3.44) & (0.88) & (1.79) & (-1.31) & (-1.96) & (5.12) & (-1.82) & (-2.32) \\
\bottomrule
\end{tabular}
\begin{flushleft}
{\footnotesize * \(p<0.10\), ** \(p<0.05\), *** \(p<0.01\). Intercepts represent risk-model-relative abnormal returns rather than CAPM alpha. QUAL = quality-minus-junk spread.}
\end{flushleft}
\end{table*}

\subsubsection{Interpretation}

The factor attribution therefore supports a consistent economic picture: AlphaZeroBeta behaves as a moderate trend follower with low beta exposure, meaningful avoidance of fast mean-reversion, and only minor quality tilts that depend on the index composition. The binary ablation in Table~\ref{tab:rl_baseline} confirms that removing the correlation penalty ($\lambda_1=0$) raises market correlations into the $0.4$--$0.6$ range and deepens drawdowns while also lowering Sharpe in every market, so the penalty acts as a beneficial regularizer rather than a pure cost to alpha at the operating point chosen here.

\subsubsection{Feature-Level Attribution}

To understand which inputs drive the learned policy, we apply two complementary attribution methods to the CNN-GRU encoder: grouped permutation importance and integrated gradients. Features are partitioned into four blocks: price/momentum signals (returns, rolling means, technical indicators), volatility and regime indicators, macro/cross-asset drivers, and sentiment/flow inputs.

For permutation importance, we shuffle each block in turn within the 2014--2024 out-of-sample window and measure the relative drop in mean Sharpe ratio (versus the unshuffled baseline), averaged across the 22 walk-forward folds and the seven markets. The marginal drops are 37\% (price/momentum), 18\% (volatility/regime), 12\% (macro/cross-asset), and 9\% (sentiment/flows); the remaining 24\% reflects interaction effects and redundancy across blocks. Integrated gradients, averaged over 256 random test-window evaluations per market, confirm that price/momentum and volatility signals dominate the policy logits, while macro and sentiment features mainly modulate position sizing.

These diagnostics are consistent with the factor loadings in Table~\ref{tab:factor_results_tstats}: AlphaZeroBeta behaves as a trend-following policy modulated by volatility-aware sizing, with significant momentum loadings and only modest tilts on the remaining factors.

\section{Discussion}\label{sec:discussion}

Across the seven equity indices in Table~\ref{tab:strategy_comparison} and the representative S\&P~500 trajectory in Figure~\ref{fig:pnl_main}, AlphaZeroBeta delivers higher Sharpe ratios than the convex baselines, smaller absolute correlations with the benchmark, and drawdowns that are uniformly shallower than buy-and-hold and shallower than the decorrelation baseline in six of seven markets (Decorr produces a slightly shallower drawdown only on 000001.SS). The same hyperparameter setting transfers across markets without per-index tuning, indicating that the multi-objective reward---rather than the choice of RL algorithm alone---is the principal driver of performance in non-stationary equity environments.

From a methodological standpoint, our findings build upon and extend earlier work on reinforcement learning in portfolio management~\cite{jiang2017, ye2020, lim2022, liang2018, yang2020, wu2021}. In contrast to approaches focused on return maximization, AlphaZeroBeta introduces a multi-objective reward function that explicitly penalizes market correlation and transaction costs. This encourages the discovery of uncorrelated alpha while promoting stable, cost-aware rebalancing.

The cumulative PnL trajectory in Figure~\ref{fig:pnl_main} shows that the AlphaZeroBeta drawdowns during the 2018Q4, 2020Q1, and 2022 episodes remain materially shallower than those of the index buy-and-hold path, consistent with the maximum drawdown column of Table~\ref{tab:strategy_comparison}.

\subsection{Baseline Behavior}

Table~\ref{tab:strategy_comparison} shows that the convex baselines trail AlphaZeroBeta in Sharpe ratio across all markets. Index B\&H is passive and inherits all of the benchmark's beta exposure; MxSharpe and Decorr, although recomputed at each rebalance, underperform for two interacting reasons. The first is estimation noise: sample means and covariances computed on short windows are unstable, so the optimizer overfits transient mean shifts and outputs concentrated portfolios that degrade once the in-sample covariance structure shifts---visible in the very low MxSharpe Sharpe on \textasciicircum{}GDAXI ($0.11$) and the consistently negative Decorr Sharpes across all markets. The second is constraint asymmetry: all three baselines enforce full investment ($\sum_i w_i = 1$) without penalizing residual beta, so any mismatch between the training and test market factor leaves them net long, compounding drawdowns during stress regimes.

AlphaZeroBeta continually re-estimates exposures through its state encoder and enforces dollar neutrality via the explicit centering-and-\(\ell_1\)-projection step in Listing~\ref{lst:alphazerobeta-env}; the correlation penalty in the reward then suppresses any residual beta arising from transaction costs or execution. This combination prevents persistent beta drift even when regimes switch, explaining why the agent maintains low correlations while delivering higher Sharpe ratios despite facing the same data.

The unconstrained RL ablation in Table~\ref{tab:rl_baseline} further isolates the contribution of the neutrality penalty. Removing the correlation term ($\lambda_1 = 0$) lets the policy lean into prevailing index trends; correlations rise into the roughly 0.4--0.6 range and maximum drawdowns deepen by approximately 7--20 percentage points relative to AlphaZeroBeta, while out-of-sample Sharpe is lower in every market under this ablation. Because the turnover penalty remains active, the ablation cannot simply neutralize exposures through rapid hedging, so regime changes translate into beta shocks. These observations confirm that AlphaZeroBeta’s superior performance does not stem solely from using DRL; it is the explicit reward shaping that enables simultaneous achievement of high Sharpe ratios and near-zero market correlation.

More broadly, the results indicate that reinforcement learning is a tractable tool for portfolio construction when the reward encodes the constraints of interest directly. Casting allocation as a constrained stochastic control problem replaces the need for prespecified factor models or a fixed mean-variance objective with policy adaptation driven by reward shaping.

\subsection{Future Directions}

The findings point toward several potential directions for future research.

\begin{enumerate}
  \item Extending the framework to multi-asset portfolios (fixed income, commodities, crypto) would allow evaluating inter-asset regime rotation and cross-asset neutrality.

  \item Risk-conditioned policies that incorporate regime-specific volatility forecasts or macroeconomic indicators could improve responsiveness to systemic shocks.

  \item Hierarchical reinforcement learning with a meta-controller would let higher-level agents allocate capital across multiple specialized AlphaZeroBeta sub-strategies.

  \item Integrating real-time deployment constraints, including execution latency and slippage modeling, would narrow the gap between backtest and live deployment.
\end{enumerate}

\section{Conclusions}\label{sec:conclusions}

This paper presents AlphaZeroBeta, a deep reinforcement learning framework for dynamic, market-neutral portfolio construction. By formulating the asset allocation problem as a Markov decision process and optimizing a composite reward that balances risk-adjusted excess return, benchmark correlation, and transaction costs, we show that reinforcement learning can support risk-aware portfolio construction beyond pure return maximization. Empirical evidence from the 2014--2024 walk-forward study supports the model's ability to maintain neutrality while delivering benchmark-relative excess returns.

Across seven global equity indices, AlphaZeroBeta achieves higher Sharpe ratios than the convex baselines and maintains near-zero benchmark correlations, with competitive drawdowns: its maximum drawdowns are uniformly shallower than the buy-and-hold path and shallower than the decorrelation baseline in six of seven markets, while the Sharpe-optimal baseline records shallower drawdowns in several markets. The model generalizes across markets with varying liquidity, structure, and volatility regimes without per-index tuning.

Future work includes extending the framework to multi-asset portfolios (fixed income, commodities, crypto), incorporating execution-latency and slippage modeling for live deployment, and exploring hierarchical formulations in which a meta-controller allocates capital across several AlphaZeroBeta sub-strategies. Macroeconomic and regime-conditioned signals are a natural way to make the reward more risk-aware in periods of structural change.

\section*{Acknowledgements}

During the preparation of this manuscript, the author used ChatGPT (OpenAI) and Claude Code (Anthropic) to assist with editing and LaTeX formatting. The author has reviewed and edited all generated content and takes full responsibility for the final version of this publication.

\backmatter%

\clearpage
\begin{appendices}

\clearpage
\section{Cumulative Performance}\label{app:charts}

This appendix reports cumulative equity curves for each benchmark so the distributional statistics in Section~\ref{sec:results} can be inspected visually. Each page covers one index and pairs a short interpretive note with the corresponding figure.

\clearpage

\noindent The SSE Composite buy-and-hold experienced multiple severe stress episodes over 2014--2024 (the 2015--2016 domestic equity crash and policy-driven declines in 2018 and 2022). AlphaZeroBeta compounds steadily through these regimes, delivering Sharpe $1.63 \pm 0.38$, maximum drawdown $-0.34 \pm 0.11$, and near-zero correlation $-0.02 \pm 0.05$ to the index (Table~\ref{tab:strategy_comparison}).
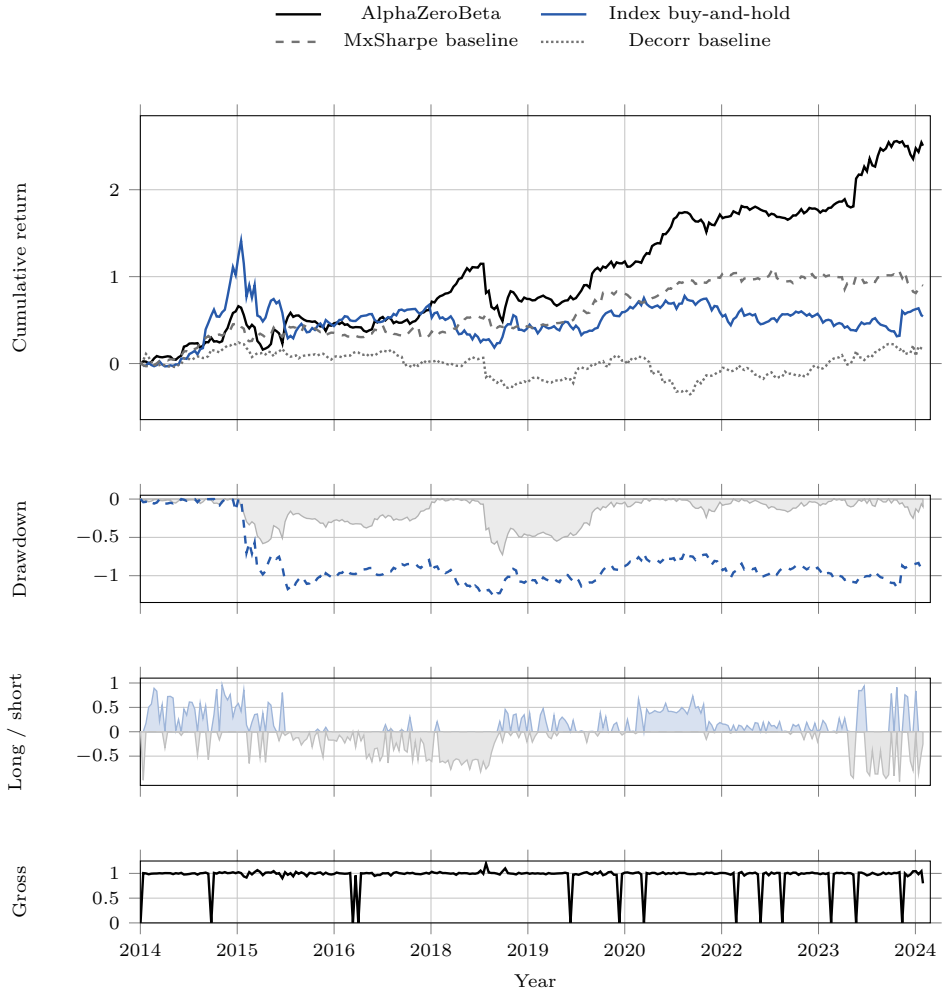
\pnlchart{figures/data/pnl_sse.csv}{figures/data/dd_sse.csv}{figures/data/exp_sse.csv}{Cumulative PnL and exposure trajectories for the SSE Composite (China, 000001.SS). Top: cumulative return; below: drawdown, long/short exposure, gross exposure.}{fig:sse}[-1.35][-1.1]

\clearpage

\noindent The DAX is a 40-stock benchmark of large German industrials and exporters, with the 2018 trade tensions, 2020 COVID drawdown, and 2022 energy crisis all visible in the index curve. AlphaZeroBeta recovers quickly from each shock and achieves the shallowest maximum drawdown across the seven markets ($-0.16 \pm 0.11$, the best DD in Table~\ref{tab:strategy_comparison}), with Sharpe $0.86 \pm 0.23$ and correlation $0.05 \pm 0.04$.
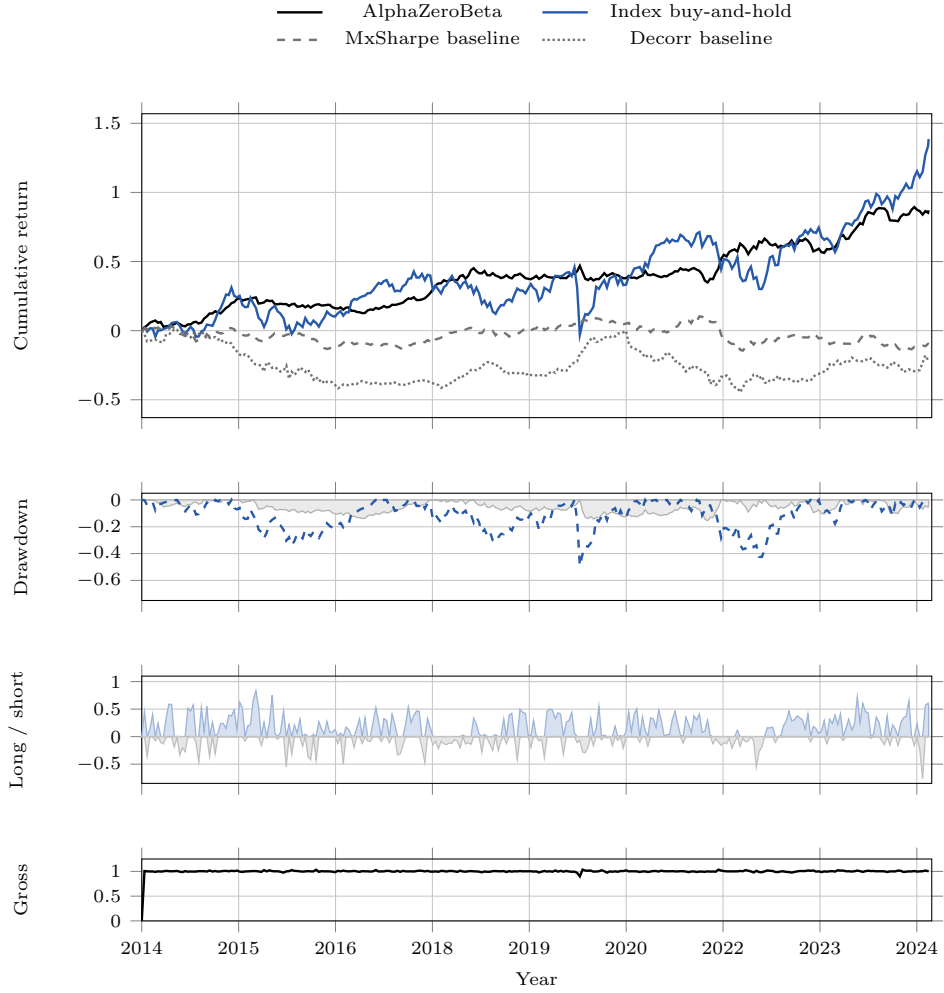
\pnlchart{figures/data/pnl_gdaxi.csv}{figures/data/dd_gdaxi.csv}{figures/data/exp_gdaxi.csv}{Cumulative PnL and exposure trajectories for the DAX (Germany, \textasciicircum{}GDAXI). Top: cumulative return; below: drawdown, long/short exposure, gross exposure.}{fig:dax}[-0.75][-0.85]

\clearpage

\noindent The Hang Seng buy-and-hold curve reflects the prolonged 2021--2023 sell-off driven by China property stress, tech-sector regulation, and capital-flow reversals. AlphaZeroBeta tolerates these abrupt volatility spikes, delivering Sharpe $1.04 \pm 0.33$, maximum drawdown $-0.21 \pm 0.22$, and near-zero correlation $0.01 \pm 0.04$ (Table~\ref{tab:strategy_comparison}).
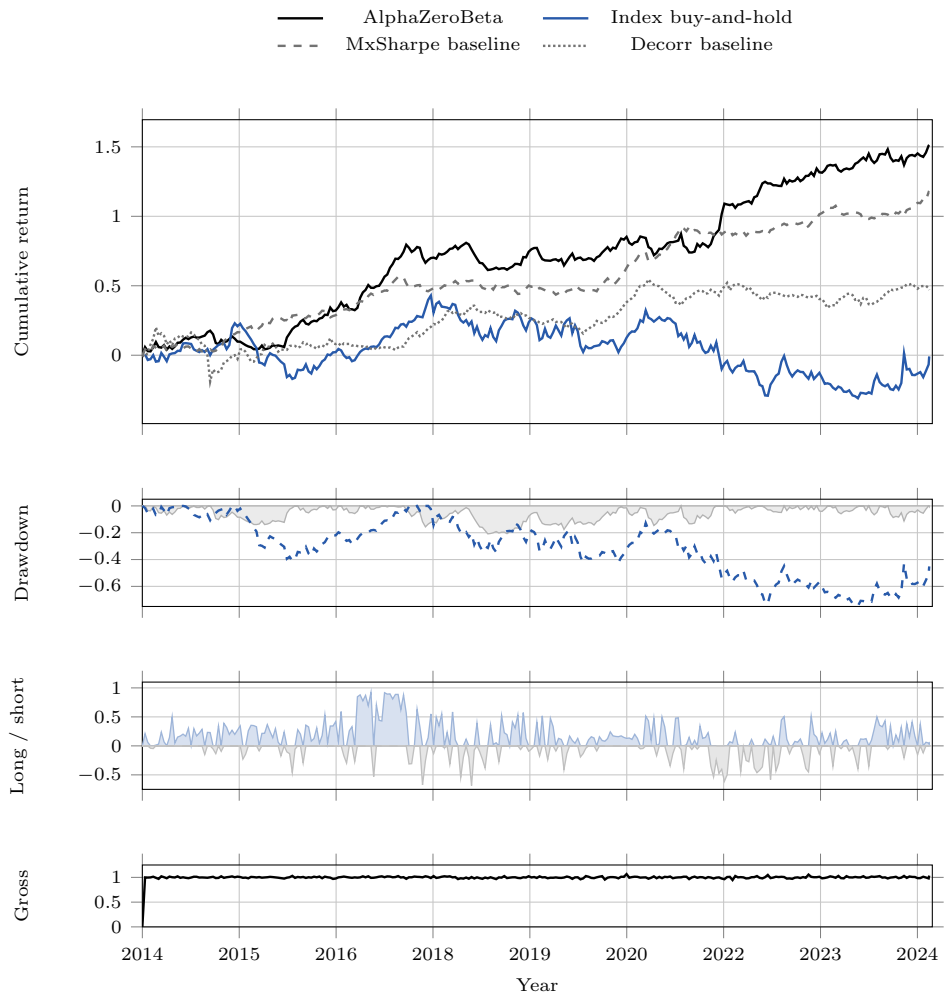
\pnlchart{figures/data/pnl_hsi.csv}{figures/data/dd_hsi.csv}{figures/data/exp_hsi.csv}{Cumulative PnL and exposure trajectories for the Hang Seng Index (Hong Kong SAR, China, \textasciicircum{}HSI). Top: cumulative return; below: drawdown, long/short exposure, gross exposure.}{fig:hsi}

\clearpage

\noindent The NASDAQ-100 had the deepest index buy-and-hold drawdown of the seven markets, driven by the 2022 rate-hiking sell-off in long-duration tech equities. AlphaZeroBeta participates in momentum phases and trims exposure when cross-sectional dispersion narrows, producing Sharpe $1.48 \pm 0.41$, maximum drawdown $-0.32 \pm 0.10$, and correlation $0.07 \pm 0.06$ (Table~\ref{tab:strategy_comparison}).
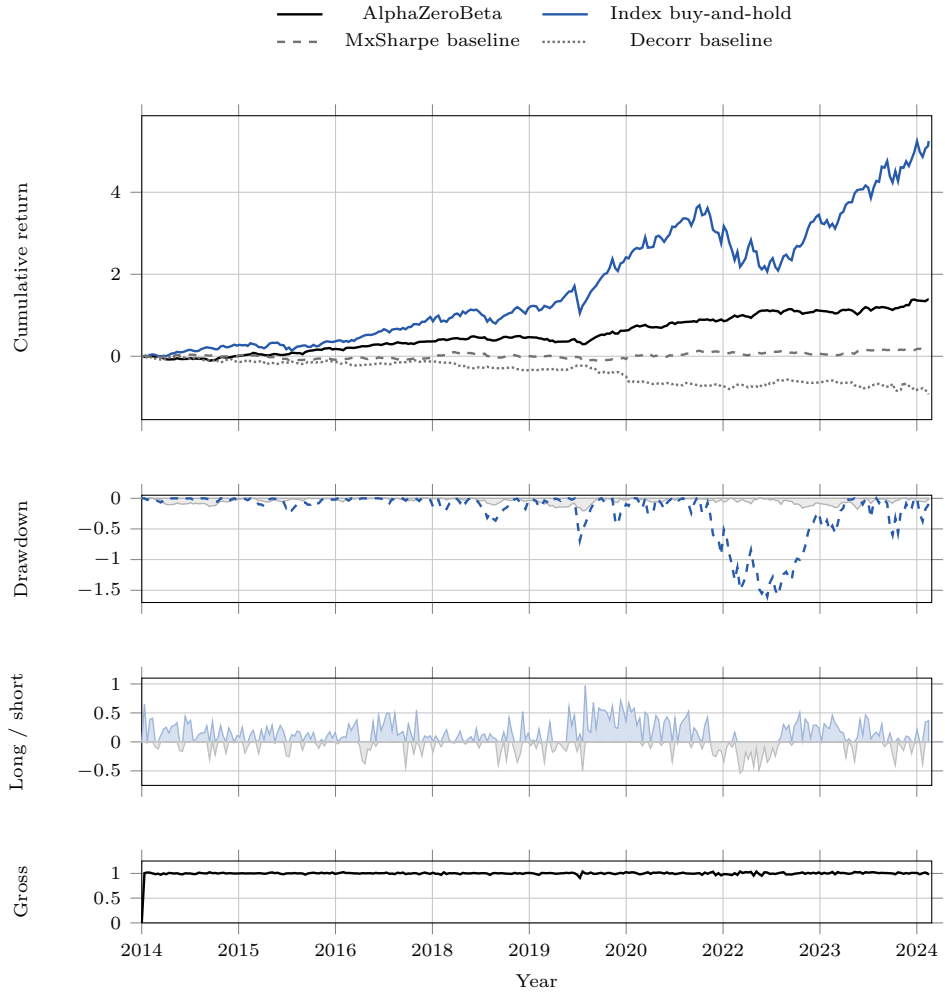
\pnlchart{figures/data/pnl_ndx.csv}{figures/data/dd_ndx.csv}{figures/data/exp_ndx.csv}{Cumulative PnL and exposure trajectories for the NASDAQ-100 (U.S., \textasciicircum{}NDX). Top: cumulative return; below: drawdown, long/short exposure, gross exposure.}{fig:ndx}[-1.7][-0.75]

\clearpage

\noindent The FTSE 100 traded sideways for much of 2014--2024, with notable shocks at the 2016 Brexit referendum, the 2020 COVID sell-off, and the 2022 mini-budget crisis. AlphaZeroBeta grows gradually in this slow regime via relative-value rotation between resource-heavy and defensive sectors: Sharpe $0.94 \pm 0.19$, maximum drawdown $-0.28 \pm 0.09$, correlation $-0.02 \pm 0.08$ (Table~\ref{tab:strategy_comparison}).
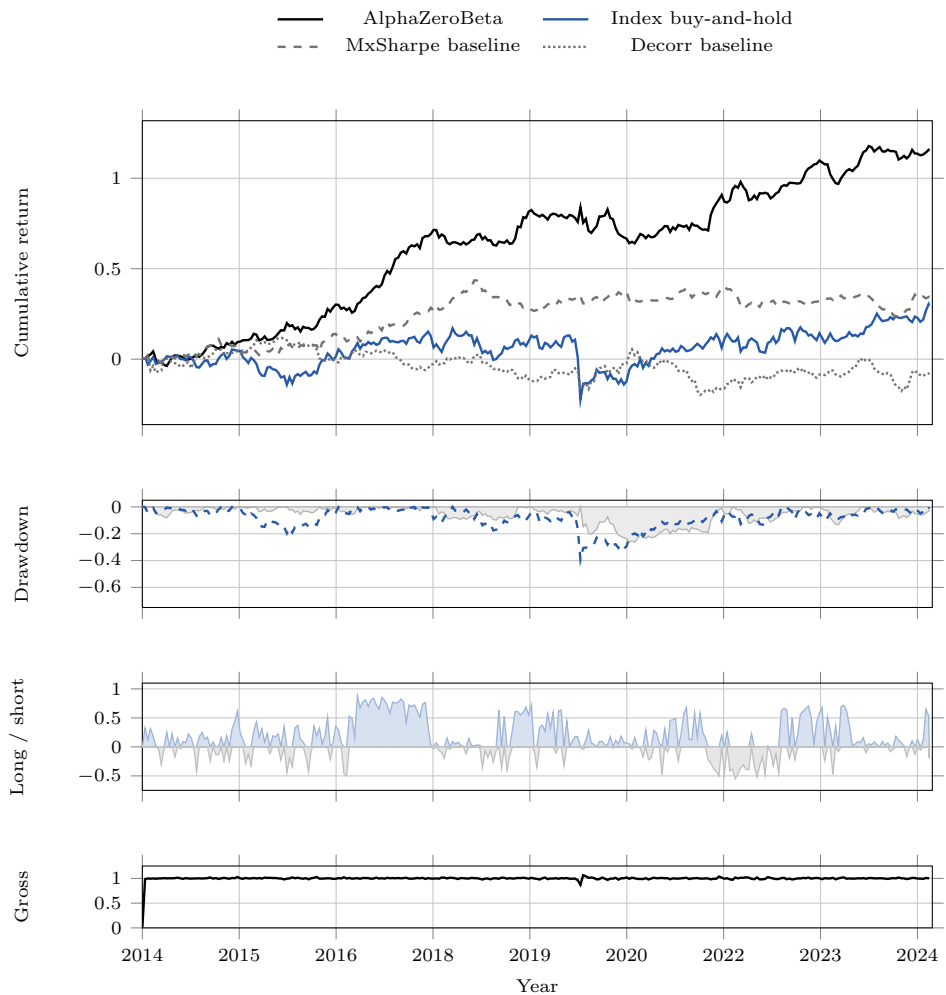
\pnlchart{figures/data/pnl_ftse.csv}{figures/data/dd_ftse.csv}{figures/data/exp_ftse.csv}{Cumulative PnL and exposure trajectories for the FTSE~100 (U.K., \textasciicircum{}FTSE). Top: cumulative return; below: drawdown, long/short exposure, gross exposure.}{fig:ftse}

\clearpage

\noindent The Dow Jones Industrial Average is a 30-stock, price-weighted index of U.S.\ blue-chip companies---the smallest universe in the panel. AlphaZeroBeta produces a smooth ascent despite the limited stock-selection breadth: Sharpe $1.20 \pm 0.28$, maximum drawdown $-0.27 \pm 0.14$, correlation $0.03 \pm 0.07$ (Table~\ref{tab:strategy_comparison}); \textasciicircum{}DJI is excluded from the factor attribution (Section~\ref{subsec:factor_exposures}) because local SMB/HML/RMW construction is statistically thin at $N{=}30$.
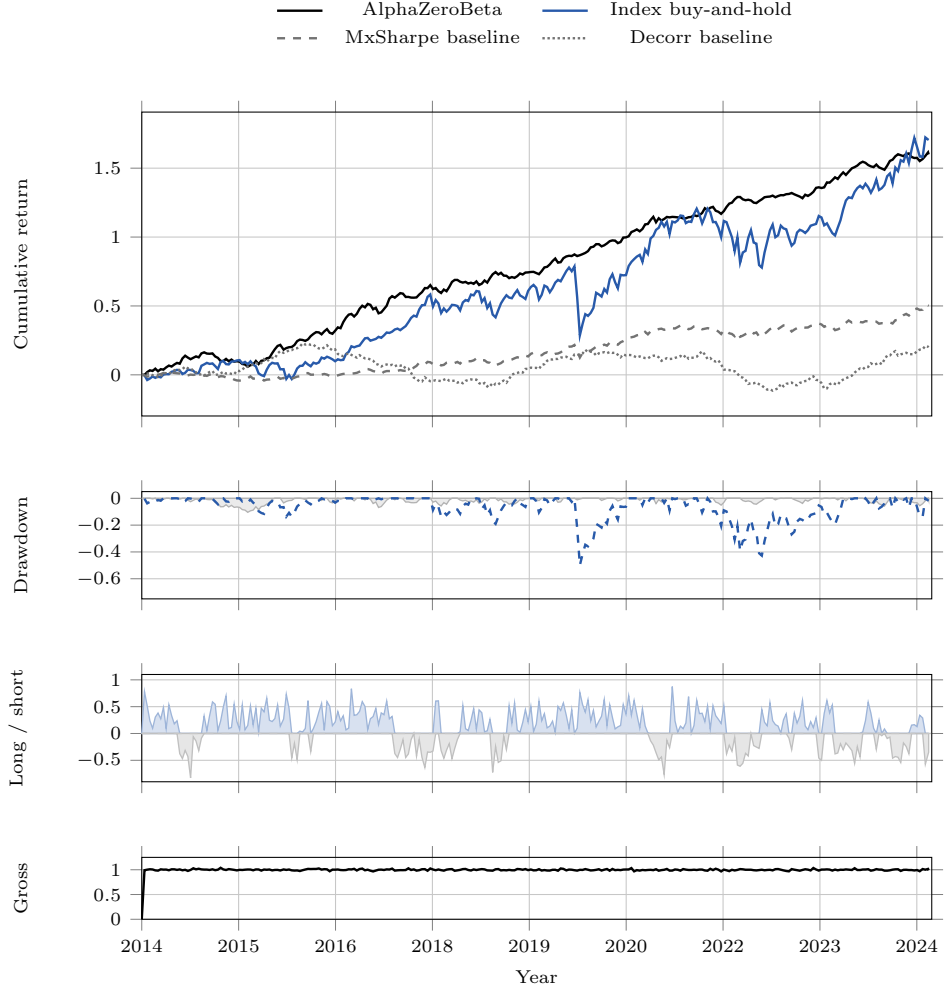
\pnlchart{figures/data/pnl_dji.csv}{figures/data/dd_dji.csv}{figures/data/exp_dji.csv}{Cumulative PnL and exposure trajectories for the Dow Jones Industrial Average (U.S., \textasciicircum{}DJI). Top: cumulative return; below: drawdown, long/short exposure, gross exposure.}{fig:dji}[-0.75][-0.9]

\clearpage

\section{Ablation Study: Return-Maximizing RL Baseline}\label{app:ablations}

To quantify the incremental benefit of an explicit neutrality penalty, we train an ablation agent that shares AlphaZeroBeta's architecture, training windows, and transaction-cost penalty but sets $\lambda_1=0$, so the reward reduces to risk-adjusted excess return minus the turnover penalty (with no correlation term). Table~\ref{tab:rl_baseline} contrasts this unconstrained RL baseline with the full AlphaZeroBeta policy across the evaluation indices. AlphaZeroBeta rows replicate the summary statistics from Table~\ref{tab:strategy_comparison} to facilitate like-for-like inspection.

\begin{table*}[t]
\centering
\caption{AlphaZeroBeta vs.\ unconstrained (return-maximizing) RL baseline. Values are mean $\pm$ standard deviation across the 22 walk-forward folds and nine independent RL initializations per fold, matching the protocol of Table~\ref{tab:strategy_comparison}; the RL baseline uses the same protocol.}\label{tab:rl_baseline}
\small
\setlength{\tabcolsep}{6pt}
\begin{tabular*}{\textwidth}{@{\extracolsep{\fill}}llccc}
\toprule
\textbf{Index} & \textbf{Method} & \textbf{Sharpe ratio} & \textbf{Max drawdown} & \textbf{Correlation} \\
\midrule
\multirow{2}{*}{000001.SS}
 & \textit{AlphaZeroBeta} & $1.63 \pm 0.38$ & $-0.34 \pm 0.11$ & $-0.02 \pm 0.05$ \\
 & RL (return-max) & $0.95 \pm 0.27$ & $-0.52 \pm 0.18$ & $0.42 \pm 0.16$ \\
\midrule
\multirow{2}{*}{\textasciicircum{}FTSE}
 & \textit{AlphaZeroBeta} & $0.94 \pm 0.19$ & $-0.28 \pm 0.09$ & $-0.02 \pm 0.08$ \\
 & RL (return-max) & $0.65 \pm 0.21$ & $-0.35 \pm 0.12$ & $0.51 \pm 0.19$ \\
\midrule
\multirow{2}{*}{\textasciicircum{}GDAXI}
 & \textit{AlphaZeroBeta} & $0.86 \pm 0.23$ & $-0.16 \pm 0.11$ & $0.05 \pm 0.04$ \\
 & RL (return-max) & $0.75 \pm 0.24$ & $-0.33 \pm 0.14$ & $0.48 \pm 0.17$ \\
\midrule
\multirow{2}{*}{\textasciicircum{}GSPC}
 & \textit{AlphaZeroBeta} & $1.61 \pm 0.48$ & $-0.26 \pm 0.15$ & $0.15 \pm 0.09$ \\
 & RL (return-max) & $1.12 \pm 0.31$ & $-0.38 \pm 0.13$ & $0.57 \pm 0.15$ \\
\midrule
\multirow{2}{*}{\textasciicircum{}HSI}
 & \textit{AlphaZeroBeta} & $1.04 \pm 0.33$ & $-0.21 \pm 0.22$ & $0.01 \pm 0.04$ \\
 & RL (return-max) & $0.58 \pm 0.25$ & $-0.41 \pm 0.16$ & $0.49 \pm 0.18$ \\
\midrule
\multirow{2}{*}{\textasciicircum{}NDX}
 & \textit{AlphaZeroBeta} & $1.48 \pm 0.41$ & $-0.32 \pm 0.10$ & $0.07 \pm 0.06$ \\
 & RL (return-max) & $1.10 \pm 0.28$ & $-0.41 \pm 0.15$ & $0.61 \pm 0.14$ \\
\midrule
\multirow{2}{*}{\textasciicircum{}DJI}
 & \textit{AlphaZeroBeta} & $1.20 \pm 0.28$ & $-0.27 \pm 0.14$ & $0.03 \pm 0.07$ \\
 & RL (return-max) & $0.88 \pm 0.26$ & $-0.36 \pm 0.12$ & $0.52 \pm 0.12$ \\
\bottomrule
\end{tabular*}
\end{table*}

The ablation confirms that, even when unconstrained RL surpasses passive benchmarks in Sharpe ratio, the absence of the correlation penalty leads to substantially larger beta exposures and weaker drawdown control. Relative to the full AlphaZeroBeta agent, the return-maximizing baseline therefore sacrifices neutrality and risk control, reinforcing the need for the composite reward.

\section{Feature Catalog}\label{app:features}
This appendix documents the full feature set used by AlphaZeroBeta. Section~\ref{app:feature-taxonomy} provides the qualitative taxonomy of Bloomberg-derived signals and their role in the model; Section~\ref{app:feature-counts} reports the approximate number of engineered signals per asset within each group together with sampling frequencies.

\subsection{Taxonomy}\label{app:feature-taxonomy}
This subsection categorizes Bloomberg-derived features used in AlphaZeroBeta’s state representation (Table~\ref{tab:bloomberg_features}). The three broad blocks—price/liquidity, fundamentals/macro, and sentiment/flow—ensure the agent observes both fast-moving market signals and slower structural drivers; these groupings complement the counts reported in Section~\ref{app:feature-counts}.

\begin{table*}[t]
\centering
\caption{Categorization of Bloomberg-derived features used in AlphaZeroBeta’s state representation.}\label{tab:bloomberg_features}
\small
\setlength{\tabcolsep}{6pt}
\renewcommand{\arraystretch}{1.1}
\begin{tabular*}{\textwidth}{@{\extracolsep{\fill}} p{0.18\textwidth} p{0.28\textwidth} p{0.48\textwidth}}
\toprule
\textbf{Category} & \textbf{Example Field / Ticker} & \textbf{Model Role} \\
\midrule
Price & \texttt{PX\_LAST} (close price) & Core input for returns and volatility. \\
Volume & \texttt{PX\_VOLUME} & Captures liquidity and trading activity. \\
Rolling Stats & rolling mean/std of \texttt{PX\_LAST} (derived) & Describes short-term distribution shifts for volatility and regime modeling. \\
Technical & \texttt{EMAVG} (EMA), \texttt{MACD}, \texttt{RSI} & Encodes momentum and trend direction. \\
Corporate Actions & \texttt{EQY\_DVD\_YLD} & Reflects dividends, buybacks, and splits. \\
Fundamentals & \texttt{RETURN\_ON\_EQUITY} & Measures profitability and efficiency. \\
Earnings & EPS surprise vs.\ estimate & Captures valuation shocks and revisions. \\
Insiders & \texttt{NUM\_INSIDER\_SHARES\_SOLD} & Signals internal sentiment and behavior. \\
Sentiment & \texttt{NEWS\_SENTIMENT\_DAILY\_MIN} & Polarity score from news headlines. \\
Governance & \texttt{ISS\_QUALITYSCORE} & ESG proxy for governance and transparency. \\
Macro & \texttt{NAPMPMI INDEX} (PMI) & Business cycle and macro regime signal. \\
Rates & \texttt{FEDL01 INDEX} (Effective Fed Funds) & Encodes monetary policy direction. \\
Volatility & \texttt{VIX INDEX} & Gauges uncertainty and hedging demand. \\
Options Flow & \texttt{12MO\_CALL\_IMP\_VOL} & Reflects speculative skew and pressure. \\
Cross-Asset & \texttt{BCOM INDEX} (Commodities) & Captures macro trends and inflation. \\
Term Structure & \texttt{USYC2Y10 INDEX} (2s10s) & Signals recession risk and yield curve slope. \\
Sectors & \texttt{XLK US EQUITY} (Tech ETF) & Used for factor and industry grouping. \\
Metadata & GICS sector / index tags & Supports stratification and normalization. \\
\bottomrule
\end{tabular*}
\end{table*}
\noindent Together these categories span the three broad blocks introduced above (price/liquidity, fundamentals/macro, and sentiment/flow) and provide both fast and slow drivers of market neutrality.

\subsection{Per-Asset Counts}\label{app:feature-counts}
Table~\ref{tab:c2-features} supplements the taxonomy by listing each feature group together with sampling frequency and the approximate number of distinct signals per asset used in our experiments. Price and liquidity features refresh daily, fundamentals and governance arrive quarterly, and macro/sentiment indicators bridge the two horizons, allowing the RL agent to balance tactical and structural inputs.

\begin{table*}[t]
\centering
\caption{Feature groups, sampling frequency, and approximate signal counts (per asset unless noted).}\label{tab:c2-features}
\small
\setlength{\tabcolsep}{6pt}
\renewcommand{\arraystretch}{1.1}
\begin{tabular*}{\textwidth}{@{\extracolsep{\fill}} p{0.16\textwidth} p{0.42\textwidth} p{0.14\textwidth} c}
\toprule
\textbf{Feature category} & \textbf{Examples} & \textbf{Frequency} & \textbf{Approx.~count} \\
\midrule
Price-based & Close, open, high, low prices; log returns; rolling means and standard deviations (5/20/60-day) & Daily & $\sim$12 \\
Volume/liquidity & Volume, turnover, bid-ask spread, rolling average volume & Daily & $\sim$6 \\
Technical indicators & EMA, MACD, RSI, Bollinger bands, momentum & Daily/weekly & $\sim$18 \\
Corporate actions & Dividend yield, split factors, shares outstanding & Daily/monthly & $\sim$4 \\
Fundamentals & EPS, ROE, book-to-market, debt-to-equity ratios & Quarterly & $\sim$24 \\
Earnings surprises & Actual vs.\ expected earnings; post-earnings drift signals & Quarterly & $\sim$8 \\
Insider trading & Net insider purchases/sales; insider sentiment indices & Daily/weekly & $\sim$4 \\
Sentiment & News polarity, social sentiment, CDS-implied risk & Daily & $\sim$6 \\
Governance & ESG/gov.\ scores, quality ratings & Quarterly & $\sim$5 \\
Macroeconomic & Inflation, unemployment, GDP, policy rates & Monthly/qtrly & $\sim$12 \\
Rates \& curves & 3M/10Y yields, yield-curve slope, term premia & Daily & $\sim$6 \\
Volatility \& options & VIX, implied vol surfaces, skew & Daily & $\sim$7 \\
Cross-asset indices & Commodity, currency, credit indices & Daily & $\sim$8 \\
Sector ETFs/indices & Sector ETF returns (e.g.\ XLK, XLF), rotation signals & Daily & $\sim$12 \\
Metadata & Market cap, GICS sector, region tags & Static & $\sim$6 \\
\bottomrule
\end{tabular*}
\end{table*}
\noindent Approximate counts refer to the number of engineered signals per asset (or per index for macro-level inputs) within each group; actual values vary slightly across universes because of data availability. This breakdown illustrates that the RL agent ingests signals ranging from daily price action to quarterly fundamentals, enabling a nuanced response to varied market regimes.

All features are normalized to zero mean and unit variance within each training batch. Missing values are forward-filled or, where appropriate, replaced by industry-median values. Lagged versions of selected variables are included to allow the agent to recognize temporal patterns. All features used in our experiments are subject to the licensing constraints discussed in Appendix~\ref{app:repro}.

\section{Reproducibility and Code Availability}\label{app:repro}
Our implementation comprises: (i) a data-acquisition layer interfacing with Bloomberg Terminal and other licensed providers; (ii) pre-processing routines that clean, winsorize, and engineer features; (iii) the RL environment and agent implementations; and (iv) evaluation scripts. Portions of (i) and specific vendor integrations in (ii) depend on proprietary APIs (e.g.\ Bloomberg Python/Java SDK and vendor-specific modules for options flow and sentiment) that cannot be redistributed. To support replication, we provide high-level descriptions and pseudocode for components independent of proprietary software. Section~\ref{app:pseudocode} presents the core pseudocode (data pre-processing, environment, agent training, and reward); Section~\ref{app:config} reports the cost schedule and key hyperparameters used in all experiments; Section~\ref{app:notes} documents implementation notes, evaluation guidance, and data licensing; Section~\ref{app:extra-code} provides additional code examples for the data-loading, environment, policy network, and recurrent PPO agent.

\subsection{Pseudocode}\label{app:pseudocode}

\subsubsection{Data pre-processing pseudocode}\label{app:preproc}

\listingcaption{lst:preproc-indices}{Daily pre-processing loop for indices.}
\begin{verbatim}
W = 20  # rolling window (business days)

for index in indices:
    prices  = load_price_data(index).sort_index()
    log_ret = np.log(prices["close"]).diff()

    # Winsorize at the [1%, 99%] PIT quantiles fitted on the 2004-2010
    # warm-up segment; thresholds are frozen before evaluation opens.
    warm_up = log_ret.loc[log_ret.index < TRAINING_START].dropna()
    lo, hi  = warm_up.quantile([0.01, 0.99]).to_numpy()
    log_ret = log_ret.clip(lower=lo, upper=hi)

    features = pd.DataFrame({
        "log_ret":        log_ret,
        "ret_mean_20":    log_ret.rolling(W).mean(),
        "ret_std_20":     log_ret.rolling(W).std(ddof=0),
        "volume_mean_20": prices["volume"].rolling(W).mean(),
    })
    features = add_technical_indicators(features, prices)  # EMA, RSI, MACD

    # Align auxiliary panels onto the price index in a single concat,
    # then forward-fill (never bfill, which would leak the future).
    df = (pd.concat([features, fundamental_df, macro_df], axis=1, join="outer")
            .reindex(prices.index).ffill())
    save_processed_data(index, df)
\end{verbatim}

The schematic single-stock feature construction in Listing~\ref{lst:preproc-single} is paired with the full production version in Listing~\ref{lst:datamodule}.

\listingcaption{lst:preproc-single}{Single-stock feature construction (schematic).}
\begin{verbatim}
def compute_features_for_stock(prices, aux, windows=(5, 20, 60)):
    log_ret = np.log(prices["close"]).diff()
    base = pd.DataFrame({"log_ret": log_ret, "vol": prices["volume"]},
                        index=prices.index)

    # Rolling mean/std on returns and volume for each window in `windows`.
    rolls = pd.concat(
        [base[col].rolling(w).agg(["mean", "std"])
                              .add_prefix(f"{col}_").add_suffix(f"_{w}")
         for col in ("log_ret", "vol") for w in windows],
        axis=1,
    )

    # Technical indicators (EMA-12/26, MACD, RSI-14, Bollinger-20).
    close = prices["close"]
    tech = pd.concat([
        ta.trend.EMAIndicator(close, 12).ema_indicator().rename("ema_12"),
        ta.trend.EMAIndicator(close, 26).ema_indicator().rename("ema_26"),
        ta.trend.MACD(close).macd_diff().rename("macd_diff"),
        ta.momentum.RSIIndicator(close, 14).rsi().rename("rsi_14"),
    ], axis=1)

    # One concat aligns every auxiliary panel onto the price grid;
    # ffill (never bfill) carries past values forward without leaking.
    auxiliary = pd.concat([df.reindex(prices.index) for df in aux.values()],
                          axis=1, keys=aux.keys())
    return (pd.concat([prices, base, rolls, tech, auxiliary], axis=1)
              .ffill()
              .dropna(how="any"))
\end{verbatim}

\subsubsection{Reinforcement-learning environment pseudocode}\label{app:env}

Listing~\ref{lst:env} sketches a single-asset toy environment used to illustrate the decision-at-$t$/execution-at-$t{+}1$ timing convention; the full cross-sectional production environment is shown in Listing~\ref{lst:alphazerobeta-env}.

\listingcaption{lst:env}{Minimal trading environment (schematic).}
\begin{verbatim}
class TradingEnv(gym.Env):
    # Decision-at-t / execution-at-t+1: action chosen at close(t) earns
    # return(t+1).
    def __init__(self, prices, features, cost_lookup, capital=1.0):
        self.prices, self.features = prices, features
        self.cost_lookup = cost_lookup       # date -> bps per side
        self.capital     = float(capital)
        self.reset()

    def reset(self):
        self.t, self.position = 0, 0          # position in {-1, 0, +1}
        self.cash = self.nav = self.prev_nav = self.capital
        return self._get_state()

    def _get_state(self):
        return np.concatenate([self.features[self.t],
                               [self.position, self.cash / self.capital]])

    def step(self, action):
        p_t, p_next = (float(self.prices["close"].iat[i])
                       for i in (self.t, self.t + 1))
        dpos = action - self.position
        tc   = self.cost_lookup(self.prices.index[self.t]) * 1e-4 * abs(dpos) * p_t
        self.cash    -= tc + dpos * p_t
        self.position = action
        self.nav      = self.cash + self.position * p_next

        # log(nav / prev_nav) is the stable form; floor avoids /0 on flat days.
        reward = float(np.log(max(self.nav, 1e-12) / max(self.prev_nav, 1e-12)))
        self.prev_nav, self.t = self.nav, self.t + 1
        done = self.t >= len(self.prices) - 1
        return self._get_state(), reward, done, {}
\end{verbatim}

\subsubsection{Agent training loop pseudocode}\label{app:agent}

\listingcaption{lst:actorcritic}{Actor-critic training loop (schematic).}
\begin{verbatim}
agent     = ActorCriticNetwork(state_dim, action_dim)
optimizer = torch.optim.Adam(agent.parameters(), lr=3e-4)

for epoch in range(num_epochs):
    state, done = env.reset(), False
    while not done:
        action, log_prob, value = agent.act(state)
        next_state, reward, done, _ = env.step(action)

        with torch.no_grad():
            next_value = agent.value(next_state) * (1.0 - float(done))
        td_target = reward + 0.99 * next_value           # gamma = 0.99
        advantage = td_target - value

        loss = (-log_prob * advantage.detach()
                + 0.5 * F.mse_loss(value, td_target))    # value-loss coef 0.5
        optimizer.zero_grad(set_to_none=True)
        loss.backward()
        optimizer.step()
        state = next_state
\end{verbatim}

\subsubsection{Reward computation}\label{app:reward}
Listing~\ref{lst:reward} implements the composite reward of Section~\ref{subsec:reward} as a pure function of the portfolio and benchmark returns at step $t$, the rolling portfolio volatility and benchmark correlation, and the pre- and post-trade weight vectors. The three terms---risk-adjusted excess return, correlation penalty, and turnover penalty---correspond to Eq.~\eqref{eq:reward_full}. Default coefficients ($\lambda_1=0.5$, $\lambda_2=0.001$) match Table~\ref{tab:hyperparams}; $\sigma_p$ is floored at $10^{-8}$ to guard the risk-adjusted term against degenerate flat-volatility windows.

\listingcaption{lst:reward}{AlphaZeroBeta reward used inside the environment.}
\begin{verbatim}
# Inputs: realised returns rp_t (portfolio), rm_t (benchmark),
# rolling volatility sigma_p_t, rolling Corr(r_p, r_m) corr_pm_t,
# pre- and post-trade weight vectors w_prev, w; coefficients lambda1, lambda2.
def alpha_zero_beta_reward(rp_t, rm_t, sigma_p_t, corr_pm_t, w, w_prev,
                           lambda1=0.5, lambda2=0.001):
    risk_adjusted = (rp_t - rm_t) / max(sigma_p_t, 1e-8)   # 1e-8 floor on sigma
    turnover      = float(np.abs(w - w_prev).sum())        # L1 weight change
    return float(risk_adjusted - lambda1 * corr_pm_t - lambda2 * turnover)
\end{verbatim}

\subsection{Configuration and Tables}\label{app:config}

\subsubsection{Execution-cost and borrow-fee schedule}\label{app:costs}
Table~\ref{tab:cost_schedule} reports the exact transaction-cost and borrow-fee constants used in all backtests. Top-decile membership is determined by trailing 60-trading-day average dollar volume within each index universe and refreshed monthly, using only information available up to each rebalance date.

\begin{table*}[t]
\centering
\caption{Deterministic execution-cost and borrow-fee schedule used in all reported backtests. Trading costs are charged per side on each rebalance; borrow fees are annualized and accrued linearly over holding days.}\label{tab:cost_schedule}
\small
\setlength{\tabcolsep}{3pt}
\begin{tabular}{l c c c p{3.2cm}}
\toprule
\textbf{Market bucket} & \textbf{\shortstack{Top-decile\\cost}} & \textbf{Other names} & \textbf{\shortstack{Borrow\\fee}} & \textbf{Applies to} \\
\midrule
U.S. large-cap & 5 bps/side & 15 bps/side & 30 bps/year & \textasciicircum{}GSPC, \textasciicircum{}NDX, \textasciicircum{}DJI \\
U.K./Germany & 10 bps/side & 15 bps/side & 45 bps/year & \textasciicircum{}FTSE, \textasciicircum{}GDAXI \\
Hong Kong SAR, China & 10 bps/side & 20 bps/side & 75 bps/year & \textasciicircum{}HSI \\
China (proxy universe) & 30 bps/side & 30 bps/side & 120 bps/year & 000001.SS \\
\bottomrule
\end{tabular}
\end{table*}

\subsubsection{Key hyperparameters}\label{app:hyperparams}

Table~\ref{tab:hyperparams} summarizes the main reinforcement learning hyperparameters used in our experiments. Where possible, we follow standard settings from widely used PPO implementations and keep these values fixed across markets; \(\lambda_1\) and \(\lambda_2\) are chosen through manual optimization on pilot experiments to balance market neutrality and turnover against Sharpe ratio, as discussed in Subsection~\ref{subsec:reward}. These values represent heuristic choices that work well across the indices tested; systematic sensitivity analysis and market-specific tuning may further improve performance and represent important directions for future research.

\begin{table*}[t]
\centering
\caption{Key reinforcement learning hyperparameters used in AlphaZeroBeta. Values refer to the main experiments; any deviations in robustness checks are stated in the text.}\label{tab:hyperparams}
\small
\setlength{\tabcolsep}{5pt}
\renewcommand{\arraystretch}{1.1}
\begin{tabular}{p{0.34\textwidth}p{0.58\textwidth}}
\toprule
\textbf{Component} & \textbf{Value / description} \\
\midrule
RL algorithm & Proximal Policy Optimization (Recurrent PPO) \\
Policy/value network & CNN-GRU encoder; each head has a 512-unit hidden ReLU layer. Value head outputs a scalar $V_\phi(s_t)$; policy head outputs an $N$-dimensional weight vector with Tanh activation, where $N$ is the universe size (30 for \textasciicircum{}DJI to ${\sim}2{,}200$ for 000001.SS) \\
Discount factor \(\gamma\) & \(0.99\) (daily data; close to unity) \\
Learning rate & \(3\times 10^{-4}\) (Adam optimizer) \\
PPO clip ratio & \(0.20\) \\
GAE parameter \(\lambda_{\mathrm{GAE}}\) & \(0.95\) \\
Entropy coefficient & \(0.01\) \\
Value loss coefficient & \(0.5\) \\
Minibatch size & \(256\) trajectories per update (across environments) \\
Training window length & 36 months (3 years) of daily data, immediately preceding the 6-month validation segment of each fold \\
Agent observation window ($n_{\mathrm{agent\_window}}$) & 100 daily timesteps per channel of the stacked daily/weekly/monthly CNN input (see Figure~\ref{fig:rl_architecture}) \\
Rolling window for $\sigma_p$ and $\mathrm{Corr}(r_p,r_m)$ & 60 business days (\(\approx\)3 months) \\
Number of PPO epochs per update & \(10\) \\
Correlation penalty \(\lambda_1\) & \(0.5\) (market-neutrality strength) \\
Turnover penalty \(\lambda_2\) & \(0.001\) (transaction-cost intensity) \\
Number of walk-forward splits \(K\) & \(22\) non-overlapping 6-month test splits (Jan~2014--Dec~2024) \\
\bottomrule
\end{tabular}
\end{table*}

\subsection{Implementation Notes and Reproducibility}\label{app:notes}

\subsubsection{Implementation notes}\label{app:impl-notes}
\begin{enumerate}
  \item Volatility \(\sigma_p(t)\) and correlation \(\mathrm{Corr}(r_p(t), r_m(t))\) are estimated over a sliding window (e.g.\ several months of daily data). Exponential weighting can improve responsiveness.
  \item Coefficients \(\lambda_1\) and \(\lambda_2\) control market neutrality and turnover penalties; optimal values depend on the asset universe and frequency.
  \item For reporting standard Sharpe ratios alongside the reward, subtract the risk-free rate from portfolio returns. The reward's $\Delta r(t) = r_p(t) - r_m(t)$ is invariant to this subtraction (the risk-free rate cancels when both legs are excess-of-RF).
\end{enumerate}

\subsubsection{Evaluation and replication}\label{app:eval}
To reproduce the paper's results, researchers require a high-quality historical dataset containing prices, volumes, corporate actions, and macro variables for the indices described in the main paper. With such data, the pre-processing pipeline and RL pseudocode above can be implemented in Python using open-source libraries such as \texttt{pandas}, \texttt{numpy}, \texttt{torch}, \texttt{gymnasium}, the \texttt{ta} technical-analysis package (matching the listings in this appendix), and \texttt{stable-baselines3}. In practice, one can rely on the recurrent extensions in \texttt{sb3-contrib} (RecurrentPPO) for a production-ready implementation of the policy gradient updates. We suggest evaluating performance with rolling walk-forward splits that keep each test window out of the corresponding training window. Key metrics include cumulative return, Sharpe ratio, maximum drawdown, and turnover.

\subsubsection{Data and licensing}\label{app:data-licensing}
The complete code base cannot be made publicly available, as certain components depend on licensed Bloomberg functionality and proprietary third-party libraries that are not eligible for redistribution. Listings~\ref{lst:preproc-indices}--\ref{lst:reward} capture the essential computations and can serve as a blueprint for independent replication.

\subsection{Additional Code Examples}\label{app:extra-code}

\listingcaption{lst:datamodule}{Encapsulating data loading and feature engineering.}
\begin{verbatim}
from dataclasses import dataclass
from functools import reduce
from pathlib import Path
import numpy as np, pandas as pd, ta
from sklearn.preprocessing import OneHotEncoder

DATA_DIR    = Path("./data")
RET_WINDOWS = (5, 20, 60)
AUX_BLOCKS  = ("fundamentals", "earnings", "insiders", "sentiment", "governance")

@dataclass(frozen=True)
class Dataset:
    daily:    np.ndarray   # (T, N, F_d) — per-ticker daily features
    weekly:   np.ndarray   # (T_w, N, F_d)
    monthly:  np.ndarray   # (T_m, N, F_d)
    returns:  np.ndarray   # (T, N) per-ticker log returns
    benchmark: np.ndarray  # (T,)    benchmark log returns
    dates:    pd.DatetimeIndex

def _read_panel(path):
    return pd.read_csv(path, parse_dates=["date"]).set_index("date")

def _technical_block(close):  # EMA-12/26, MACD, RSI-14, Bollinger-20
    macd = ta.trend.MACD(close)
    boll = ta.volatility.BollingerBands(close, 20, 2.0)
    return pd.concat([
        ta.trend.EMAIndicator(close, 12).ema_indicator().rename("ema_12"),
        ta.trend.EMAIndicator(close, 26).ema_indicator().rename("ema_26"),
        macd.macd().rename("macd"),       macd.macd_signal().rename("macd_signal"),
        macd.macd_diff().rename("macd_diff"),
        ta.momentum.RSIIndicator(close, 14).rsi().rename("rsi_14"),
        boll.bollinger_lband().rename("bb_low"),
        boll.bollinger_hband().rename("bb_high"),
    ], axis=1)

class DataModule:
    """Builds aligned daily/weekly/monthly tensors for the RL agent.
    Panels are reindexed onto the common daily grid and ffill'd (never bfill)."""

    def __init__(self, stock_list, benchmark_ticker):
        self.stock_list, self.benchmark_ticker = stock_list, benchmark_ticker
        meta = pd.read_csv(DATA_DIR / "stock_metadata.csv")
        ohe  = OneHotEncoder(sparse_output=False).fit(meta[["sector"]])
        self.sector_ohe = pd.DataFrame(
            ohe.transform(meta[["sector"]]),
            index=meta["ticker"].values,
            columns=ohe.get_feature_names_out(["sector"]),
        )

    def _compute_single_stock_features(self, ticker):
        prices  = _read_panel(DATA_DIR / f"{ticker}.csv")
        log_ret = np.log(prices["close"]).diff()
        vol     = prices["volume"]
        rolls = pd.concat([pd.DataFrame({
            f"ret_mean_{w}": log_ret.rolling(w).mean(),
            f"ret_std_{w}":  log_ret.rolling(w).std(ddof=0),
            f"vol_mean_{w}": vol.rolling(w).mean(),
            f"vol_std_{w}":  vol.rolling(w).std(ddof=0),
        }) for w in RET_WINDOWS], axis=1)
        aux = [_read_panel(DATA_DIR / f"{ticker}_{name}.csv") for name in AUX_BLOCKS]
        # One concat aligns everything on the price index; ffill (never bfill).
        return pd.concat([prices.assign(log_ret=log_ret), rolls,
                          _technical_block(prices["close"]), *aux],
                         axis=1).ffill()

    def build_dataset(self):
        # 1) Per-ticker daily features with broadcast sector one-hot.
        raw = {}
        for ticker in self.stock_list:
            f = self._compute_single_stock_features(ticker)
            sector = self.sector_ohe.loc[ticker]
            f[sector.index] = sector.values                  # broadcast row
            raw[ticker] = f.dropna(how="any")

        # 2) Common date grid via Index reduction (O(N), preserves DatetimeIndex).
        common = reduce(pd.Index.intersection,
                        (df.index for df in raw.values())).sort_values()
        feature_cols = raw[self.stock_list[0]].columns.drop("log_ret")
        stack = lambda key: np.stack(
            [raw[t].loc[common, key].to_numpy() for t in self.stock_list], axis=1)
        daily, returns = stack(feature_cols), stack("log_ret")

        # 3) Weekly/monthly aggregations (last value in each bucket).
        def _resample(rule):
            panels = {t: df.resample(rule).last() for t, df in raw.items()}
            grid = reduce(pd.Index.intersection,
                          (p.index for p in panels.values())).sort_values()
            return np.stack([panels[t].loc[grid, feature_cols].to_numpy()
                             for t in self.stock_list], axis=1)
        weekly, monthly = _resample("W-FRI"), _resample("ME")  # PIT-safe rules

        # 4) Benchmark on the common daily grid.
        bench = _read_panel(DATA_DIR / f"{self.benchmark_ticker}.csv")
        bench_ret = np.log(bench["close"]).diff().reindex(common).to_numpy()
        return Dataset(daily, weekly, monthly, returns, bench_ret, common)
\end{verbatim}

\listingcaption{lst:alphazerobeta-env}{Market-neutral environment with custom reward.}
\begin{verbatim}
import gym
import numpy as np

# Cross-sectional market-neutral environment. State: stacked daily/weekly/
# monthly feature tensors + previous weights. Action: target weights in
# [-1, 1]^N, projected to dollar-neutral with L1-budget <= 1.
# Reward: composite (risk-adjusted excess return - lambda_corr * corr
# - lambda_cost * turnover); see Section 3 reward function for full form.
class AlphaZeroBetaEnv(gym.Env):
    def __init__(self, daily, weekly, monthly, returns, benchmark,
                 lambda_corr=0.5, lambda_cost=0.001, vol_window=60):
        super().__init__()
        self.daily, self.weekly, self.monthly = daily, weekly, monthly
        self.returns, self.benchmark = returns, benchmark
        self.T, self.N, self.F_d     = daily.shape
        self.lambda_corr, self.lambda_cost, self.vol_window = (
            lambda_corr, lambda_cost, vol_window)
        self.action_space = gym.spaces.Box(-1, 1, shape=(self.N,), dtype=np.float32)
        obs_dim = self.N * (self.F_d + weekly.shape[2] + monthly.shape[2]) + self.N
        self.observation_space = gym.spaces.Box(-np.inf, np.inf, shape=(obs_dim,),
                                                dtype=np.float32)

    def reset(self):
        self.t, self.prev_weights = 0, np.zeros(self.N, dtype=np.float32)
        return self._get_obs()

    def _get_obs(self):
        # Use the most recent CLOSED weekly/monthly bucket: (t//k)-1 skips
        # the in-progress bucket whose end-of-period values are unobservable.
        idx = lambda k, n: int(np.clip(self.t // k - 1, 0, n - 1))
        return np.concatenate([
            self.daily[self.t].ravel(),
            self.weekly[idx(5,  self.weekly.shape[0])].ravel(),
            self.monthly[idx(21, self.monthly.shape[0])].ravel(),
            self.prev_weights,
        ])

    def step(self, action):
        weights = action - action.mean()                  # dollar-neutral
        gross = np.abs(weights).sum()
        if gross > 1.0:
            weights = weights / gross                     # ||w||_1 <= 1
        weights = weights.astype(np.float32)

        # Decision-at-t, execution-at-t+1.
        rp_t = float(weights @ self.returns[self.t + 1])
        rm_t = float(self.benchmark[self.t + 1])

        # Rolling sigma_p and Corr over weights HELD on [t-W, t), so r_p(s)
        # is realised (not in-sample). Empty on the very first step.
        if self.t == 0:
            sigma_p, corr = 1e-8, 0.0
        else:
            s = max(0, self.t - self.vol_window)
            rp_hist = self.returns[s:self.t] @ self.prev_weights
            rm_hist = self.benchmark[s:self.t]
            sigma_p = max(float(rp_hist.std(ddof=0)), 1e-8)
            corr    = (float(np.corrcoef(rp_hist, rm_hist)[0, 1])
                       if rp_hist.size > 1 else 0.0)

        turnover = float(np.abs(weights - self.prev_weights).sum())
        reward   = ((rp_t - rm_t) / sigma_p
                    - self.lambda_corr * corr - self.lambda_cost * turnover)
        self.prev_weights, self.t = weights, self.t + 1
        done = self.t >= self.T - 2      # need returns[t+1] to be valid
        obs  = self._get_obs() if not done else None
        return obs, reward, done, {}
\end{verbatim}

\listingcaption{lst:policy-net}{Multi-scale policy network (CNN-GRU with heads).}
\begin{verbatim}
import torch
import torch.nn as nn

# 1-D CNN over (daily, weekly, monthly) channels followed by a GRU.
# Stacking the three resolutions on the channel axis lets one conv stack
# mix horizons before the recurrent layer (kernels 8/4/3, strides 4/2/1,
# filters 32/64/64).
class MultiScaleEncoder(nn.Module):
    def __init__(self, feature_dim, hidden_size=512, agent_window=100):
        super().__init__()
        in_ch  = 3 * feature_dim
        layers, prev = [], in_ch
        for out_ch, k, s in zip((32, 64, 64), (8, 4, 3), (4, 2, 1)):
            layers += [nn.Conv1d(prev, out_ch, kernel_size=k, stride=s),
                       nn.ReLU(inplace=True)]
            prev = out_ch
        self.conv = nn.Sequential(*layers, nn.Flatten())
        with torch.no_grad():
            self.flat_dim = self.conv(torch.zeros(1, in_ch, agent_window)).numel()
        self.gru = nn.GRU(self.flat_dim, hidden_size, batch_first=True)

    def forward(self, daily, weekly, monthly, hidden):
        feat = self.conv(torch.cat([daily, weekly, monthly], dim=1)).unsqueeze(1)
        out, new_hidden = self.gru(feat, hidden)
        return out.squeeze(1), new_hidden

class PolicyValueNet(nn.Module):
    """Shared encoder + Tanh-bounded policy head (N assets) + scalar value head."""

    def __init__(self, feature_dim, num_assets, hidden_size=512, head_hidden=512):
        super().__init__()
        self.encoder     = MultiScaleEncoder(feature_dim, hidden_size)
        self.policy_head = nn.Sequential(
            nn.Linear(hidden_size, head_hidden), nn.ReLU(inplace=True),
            nn.Linear(head_hidden, num_assets), nn.Tanh())
        self.value_head  = nn.Sequential(
            nn.Linear(hidden_size, head_hidden), nn.ReLU(inplace=True),
            nn.Linear(head_hidden, 1))

    def forward(self, daily, weekly, monthly, hidden):
        embedding, new_hidden = self.encoder(daily, weekly, monthly, hidden)
        return (self.policy_head(embedding),
                self.value_head(embedding).squeeze(-1),
                new_hidden)
\end{verbatim}

\listingcaption{lst:rppo}{Agent and training with recurrent PPO (skeleton).}
\begin{verbatim}
import logging
from dataclasses import dataclass
import numpy as np, torch
import torch.nn.functional as F
from torch.optim import Adam

GAMMA, GAE_LAMBDA, PPO_CLIP, PPO_EPOCHS = 0.99, 0.95, 0.20, 10
log = logging.getLogger("alphazerobeta")

@dataclass
class Rollout:
    obs_seq:   list           # list of (daily, weekly, monthly) tensor triples
    actions:   torch.Tensor   # untransformed samples
    rewards:   torch.Tensor
    values:    torch.Tensor
    log_probs: torch.Tensor

class AlphaZeroBetaAgent:
    """Recurrent PPO agent for the AlphaZeroBetaEnv."""

    def __init__(self, env, feature_dim, hidden_size=512, lr=3e-4):
        self.env         = env
        self.model       = PolicyValueNet(feature_dim, env.N, hidden_size)
        self.optimizer   = Adam(self.model.parameters(), lr=lr)
        self.hidden_size = hidden_size

    def _slice(self, t):
        env = self.env
        idx = lambda k, n: int(np.clip(t // k - 1, 0, n - 1))
        as_t = lambda a: torch.as_tensor(a, dtype=torch.float32).unsqueeze(0)
        return (as_t(env.daily[t]),
                as_t(env.weekly[idx(5,  env.weekly.shape[0])]),
                as_t(env.monthly[idx(21, env.monthly.shape[0])]))

    def collect_trajectory(self, horizon=200):
        obs_seq, acts, rewards, values, log_probs = [], [], [], [], []
        hidden = torch.zeros(1, 1, self.hidden_size)
        self.env.reset()
        for _ in range(horizon):
            inputs = self._slice(self.env.t)
            mean, value, hidden = self.model(*inputs, hidden)
            dist     = torch.distributions.Normal(mean, scale=0.1)
            sample   = dist.sample()
            log_prob = dist.log_prob(sample).sum(dim=-1)
            action   = torch.tanh(sample).detach().cpu().numpy()[0]
            _, reward, done, _ = self.env.step(action)
            obs_seq.append(inputs)
            acts.append(sample.squeeze(0))
            rewards.append(reward)
            values.append(value.squeeze(0))
            log_probs.append(log_prob.squeeze(0))
            if done: break
        return Rollout(obs_seq, torch.stack(acts),
                       torch.tensor(rewards, dtype=torch.float32),
                       torch.stack(values).detach(),
                       torch.stack(log_probs).detach())

    def _gae(self, rewards, values):
        # GAE (Schulman et al., 2016); bootstrap V(s_{T+1}) = 0 at horizon.
        advantages = torch.zeros_like(rewards)
        gae = 0.0
        for t in reversed(range(len(rewards))):
            v_next = values[t + 1] if t + 1 < len(values) else 0.0
            delta  = rewards[t] + GAMMA * v_next - values[t]
            gae    = delta + GAMMA * GAE_LAMBDA * gae
            advantages[t] = gae
        returns = advantages + values
        advantages = (advantages - advantages.mean()) / (advantages.std() + 1e-8)
        return advantages, returns

    def ppo_update(self, rollout):
        advantages, returns = self._gae(rollout.rewards, rollout.values)
        for _ in range(PPO_EPOCHS):
            hidden = torch.zeros(1, 1, self.hidden_size)
            new_log_probs, new_values = [], []
            for k, inputs in enumerate(rollout.obs_seq):
                mean, value, hidden = self.model(*inputs, hidden)
                dist = torch.distributions.Normal(mean, scale=0.1)
                new_log_probs.append(
                    dist.log_prob(rollout.actions[k]).sum(-1).squeeze())
                new_values.append(value.squeeze(0))
            new_log_probs = torch.stack(new_log_probs)
            new_values    = torch.stack(new_values)

            ratio   = (new_log_probs - rollout.log_probs).exp()
            clipped = torch.clamp(ratio, 1 - PPO_CLIP, 1 + PPO_CLIP)
            loss = (-torch.min(ratio * advantages, clipped * advantages).mean()
                    + 0.5 * F.mse_loss(new_values, returns))
            self.optimizer.zero_grad(set_to_none=True)
            loss.backward()
            torch.nn.utils.clip_grad_norm_(self.model.parameters(), 0.5)
            self.optimizer.step()

    def train(self, iterations=10):
        for it in range(iterations):
            rollout = self.collect_trajectory()
            self.ppo_update(rollout)
            log.info("iter=%d  steps=%d  reward_mean=%.4f",
                     it, len(rollout.rewards), float(rollout.rewards.mean()))
\end{verbatim}


\end{appendices}

\bibliographystyle{sn-bibliography}

\end{document}